\documentclass[]{spie}

\usepackage{amsmath,amsfonts,amssymb}
\usepackage{graphicx}
\usepackage{subcaption}
\usepackage{todonotes}
\usepackage[colorlinks=true, allcolors=blue]{hyperref}

\title{Studies of Systematic Uncertainties for Simons Observatory: Optical Effects and Sensitivity Considerations}

\author[a]{Patricio A. Gallardo}
\author[b]{Jon Gudmundsson}
\author[a]{Brian J. Koopman}
\author[c]{Frederick T. Matsuda}
\author[d]{Sara M. Simon}
\affil[a]{Department of Physics, Cornell University, Ithaca, NY 14853, USA}
\affil[b]{The Oskar Klein Centre for Cosmoparticle Physics, Department of Physics, Stockholm University, AlbaNova, SE-106 91 Stockholm, Sweden}
\affil[c]{Kavli Institute for the Physics and Mathematics of the Universe (WPI), The University of Tokyo Institutes of Advanced Study, The University of Tokyo, Kashiwa, Chiba, 277-8583, Japan}
\affil[d]{Department of Physics, University of Michigan, Ann Arbor, MI 48103, USA}

\author[e]{Aamir Ali}
\author[f]{Sean Bryan}
\author[e]{Yuji Chinone}
\author[g]{Gabriele Coppi}
\author[h]{Nicholas Cothard}
\author[i]{Mark J. Devlin}
\author[g]{Simon Dicker}
\author[l]{Giulio Fabbian}
\author[m]{Nicholas Galitzki}
\author[e, n]{Charles A. Hill}
\author[m]{Brian Keating}
\author[n, p]{Akito Kusaka}
\author[q]{Jacob Lashner}
\author[e, n, r]{Adrian T. Lee}
\author[g]{Michele Limon}
\author[s]{Philip D. Mauskopf}
\author[d]{Jeff McMahon}
\author[t]{Federico Nati}
\author[a]{Michael D. Niemack}
\author[i]{John L. Orlowski-Scherer}
\author[j]{Stephen C. Parshley}
\author[u]{Giuseppe Puglisi}
\author[v]{Christian L Reichardt}
\author[w]{Maria Salatino}
\author[k]{Suzanne Staggs}
\author[x]{Aritoki Suzuki}
\author[a]{Eve M. Vavagiakis}
\author[y]{Edward J. Wollack}
\author[g]{Zhilei Xu}
\author[g]{Ningfeng Zhu}

\affil[e]{Department of Physics, University of California, Berkeley, Berkeley, CA, USA}
\affil[f]{School of Electrical, Computer and Energy Engineering, Arizona State University, Tempe, AZ, USA}
\affil[g]{Department of Physics \& Astronomy, University of Pennsylvania, Philadelphia, Pennsylvania, PA, USA}
\affil[h]{Department of Applied and Engineering Physics, Cornell University, Ithaca, NY 14853, USA}
\affil[i]{Department of Physics, University of Pennsylvania, Philadelphia, Pennsylvania, PA, USA}
\affil[j]{Department of Astronomy, Cornell University, Ithaca, NY 14853, USA}
\affil[k]{Department of Physics, Princeton University, Princeton, NJ, USA}
\affil[l]{Institut d' Astrophysique Spatiale, CNRS (UMR 8617), Univ. Paris-Sud, Universit\'e Paris-Saclay, B\^at.121, 91405 Orsay, France}
\affil[m]{Department of Physics University of California San Diego, La Jolla CA, USA}
\affil[n]{Physics Division, Lawrence Berkeley National Laboratory, Berkeley, USA}
\affil[o]{Physics Division, Lawrence Berkeley National Laboratory, Berkeley, CA, USA}
\affil[p]{Department of Physics, The University of Tokyo, Tokyo, Japan}
\affil[q]{Department of Physics, University of Southern California, Los Angeles, CA, USA}
\affil[r]{Radio Astronomy Laboratory, University of California, Berkeley, Berkeley, CA 92093, USA}
\affil[s]{School of Earth and Space Exploration and Department of Physics, Arizona State University}
\affil[t]{University of Pennsylvania, Philadelphia, Pennsylvania, PA, USA}
\affil[u]{Department of Physics, Stanford University, Stanford, California, CA 94305}
\affil[v]{School of Physics, University of Melbourne, Melbourne, Australia}
\affil[w]{AstroParticle and Cosmology, (APC) laboratory, Paris Diderot University, Paris, 75013, France}
\affil[x]{Physics Division, Lawrence Berkeley National Laboratory, Berkeley, USA}
\affil[y]{NASA/Goddard Space Flight Center, Greenbelt, MD, USA}

\authorinfo{Further author information: (Send correspondence to P.A.G.)\\P.A.G.: E-mail: pag227@cornell.edu}

\usepackage{aas_macros}
 
\begin{document} 
\maketitle

\begin{abstract}
The Simons Observatory (SO) is a new experiment that aims to measure the cosmic microwave background (CMB) in temperature and polarization.  SO will measure the polarized sky over a large range of microwave frequencies and angular scales using a combination of small ($\sim$0.5~m) and large ($\sim$6~m) aperture telescopes and will be located in the Atacama Desert in Chile. This work is part of a series of papers studying calibration, sensitivity, and systematic errors for SO. In this paper, we discuss current efforts to model optical systematic effects, how these have been used to guide the design of the SO instrument, and how these studies can be used to inform  instrument design of future experiments like CMB-S4. While optical systematics studies are underway for both the small aperture and large aperture telescopes, we limit the focus of this paper to the more mature large aperture telescope design for which our studies include: pointing errors, optical distortions, beam ellipticity, cross-polar response, instrumental polarization rotation and various forms of sidelobe pickup.
\end{abstract}


\section{Introduction}

\label{sec:intro}

Measurements of the cosmic microwave background (CMB) at large\cite{ahmed_bicep3:_2014} (degree) and small\cite{thornton_atacama_2016,benson2014spt,suzuki2016polarbear} (arcminute) angular scales have been used in recent decades to constrain parameters of the $\Lambda$CDM standard model of cosmology. These measurements can be used to probe inflation\cite{planck_collaboration_planck_2018}, detect galaxy clusters\cite{hasselfield_atacama_2013}, measure cluster dynamics\cite{hand_evidence_2012}, and constrain the sum of the neutrino masses among other scientific goals.

To detect the faint light from the CMB, experiments require high sensitivity and control over image fidelity (systematics). Mechanisms such as Thompson scattering and gravitational lensing produce polarized signals that motivate polarization-sensitive experiments. Separation of CMB signals from foregrounds requires multi-frequency observations. Improving uncertainties on cosmological parameters motivates the need for large-area surveys, which also benefits cluster science and cross-correlation studies with optical surveys. 

Simons Observatory (SO) is an upcoming experiment that will measure the CMB in both temperature and polarization across a wide range of frequencies (27--270~GHz) and angular scales (degree to arcminute) from the Atacama Desert in Chile. SO will achieve this through the use of both a large aperture reflective cross-Dragone telescope with a 6~m primary aperture (in the same configuration as the CCAT-prime telescope\cite{parshley_ccat-prime:_2018}) and multiple 0.5~m aperture refractive telescopes. A description of the SO large aperture telescope (LAT) is given in Galitzki et al. 2018 \cite{Galitzki2018}. In this paper, we discuss optical systematic studies for SO and how these studies are being used to provide design feedback to minimize systematics in the observations. We model many of the optical systematics in both sets of telescopes using similar methods. We limit our focus to the optical systematic studies on the LAT. In this work we will discuss physical and phenomenological models that have been built and validated in previous CMB experiments. A description of the SO instrument optical design is given in Dicker et al. 2018\cite{Dicker18}. In Sec.~\ref{sec:opticaldistortions} we discuss optical distortions, cross-polar response is discussed in Sec.~\ref{polSyst},  in Sec.~\ref{pointing} we discuss pointing errors and various forms of sidelobe pickup are discussed in Sec.~\ref{sec:sidelobepickup}. This paper is part of a series of papers on the systematic and calibration studies for SO~\cite{crowley18,salatino18,bryan18}. We are combining the detailed results of the full SO systematics and calibration studies into a comprehensive study that will be released in a series of future papers to the community for use in developing future experiments like CMB-S4.

\section{Optical distortions}
	\label{sec:opticaldistortions}
The initial LAT receiver (LATR) deployment will consist of seven optics tubes located at the focal point of the two mirror system\cite{Zhu2018,Dicker18}. These seven optics tubes are arranged as a central tube with an hexagonal pattern around it. Space for another ring of six additional optics tubes is incorporated into this design to allow for future expansion of the instrument. Each optics tube consists of three lenses and a Lyot stop. Infrared blocking and band selection is achieved with stacks of filters placed inside the cryogenic optics tube. The optical design is described in Dicker et al 2018. \cite{Dicker18}

We quantify optical distortions by evaluating the Strehl ratio of the optical system, including mirrors and the three-lens re-imaging optics contained in the optics tubes. The Strehl ratio (eq.~\ref{eq:strehl}) is the measure of the ratio of the peak intensity of the diffraction point spread function (PSF) to the peak of the intensity of the diffraction PSF in the absence of aberrations~\cite{zemax_llc_zemax_2016}. In this calculation, diffraction is computed using a scalar theory of diffraction as it is common in current ray-tracing packages \cite{zemax_llc_zemax_2016}.

\begin{equation}
\label{eq:strehl}
S = \frac{I_{\rm PSF} ^{\rm max}}{I^{\rm max}_{\rm No\,Aberration\,PSF}}.
\end{equation}

Our studies evaluate the Strehl ratio $S$ at $\lambda=1.1$~mm (the shortest wavelength of operation) and compute the percentage of the focal plane where $S$ is  greater than 0.8 (where we define diffraction limited performance). This area coverage is used to compare designs and evaluate performance among optics tubes.

\begin{figure}
\centering
\includegraphics[width=\textwidth]{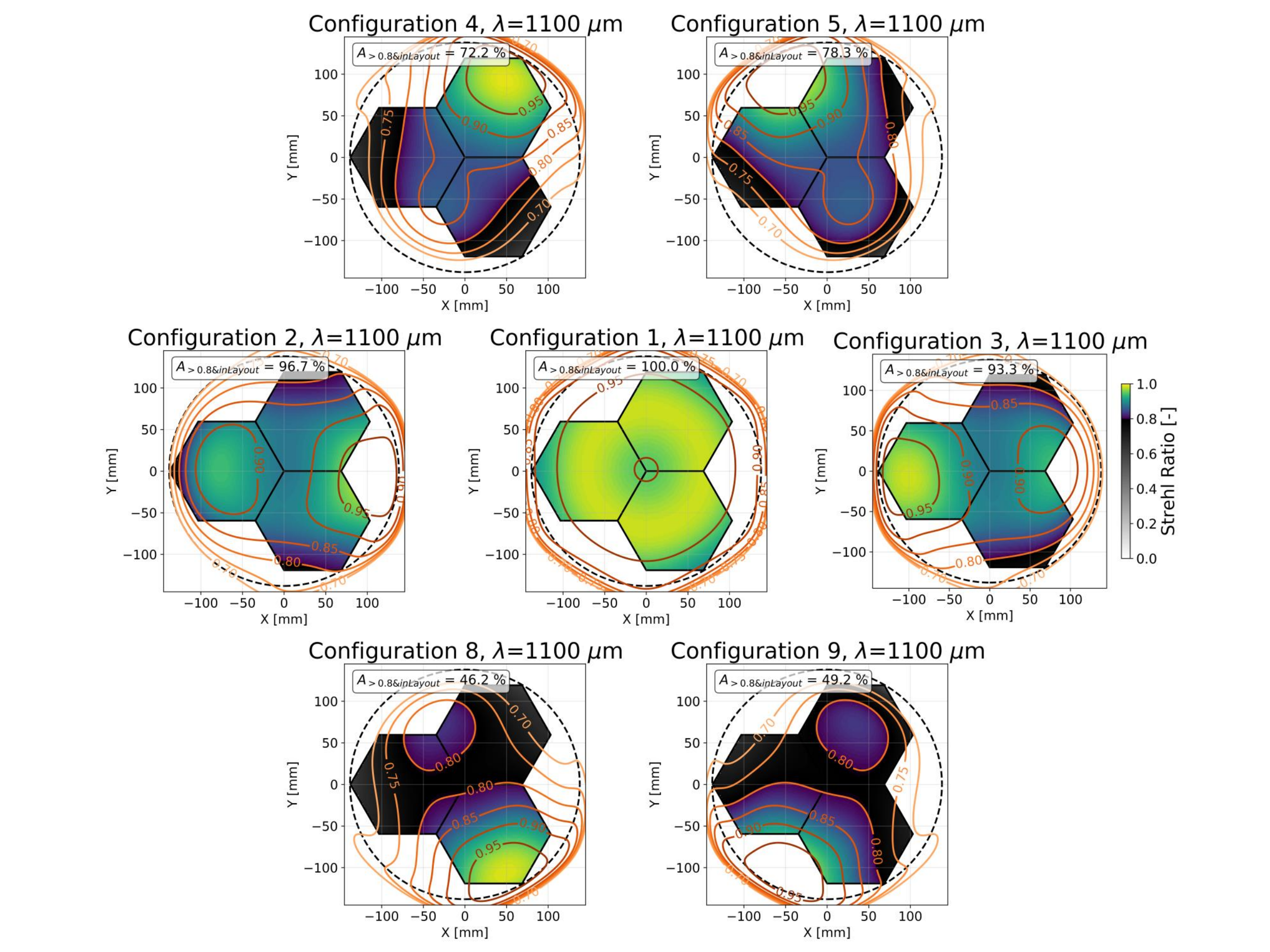}
\caption{Strehl ratios of the inner seven optics tubes of one preliminary optical design at 1.1~mm wavelength. Area shown corresponds to the fraction of the area with Strehl ratio greater than 0.8 and within a tiled hexagonal pattern at the focal plane of the system. Note that there is still one rotational degree of freedom where the angle of the tiled pattern at the focal plane is varied to allow maximum coverage. This will be explored in future work.}
\label{fig:imgqual_plot}
\end{figure}

After optimization, we find that the center tube has the best image quality coverage, with 100\% of the array above the $S=0.8$ threshold. Area above the $S=0.8$ limit ranges from $\sim 50\%$ to 85\% across optics tubes. Figure \ref{fig:imgqual_plot} summarizes these results. For simplicity, the rotational degree of freedom, where the detector hexagonal tiles are rotated to maximize the covered area is not explored and is left for future work. The optical design was optimized without taking this into account, and the hexagonal pattern was evaluated after optimization. Note that for wavelengths longer than 1.1~mm, the larger beam size increases Strehl ratios and the area above $S=0.8$ is 100\% in all seven tubes.

\subsection{Beam FWHM}    
\begin{figure}
\centering
\includegraphics[width=0.9\textwidth]{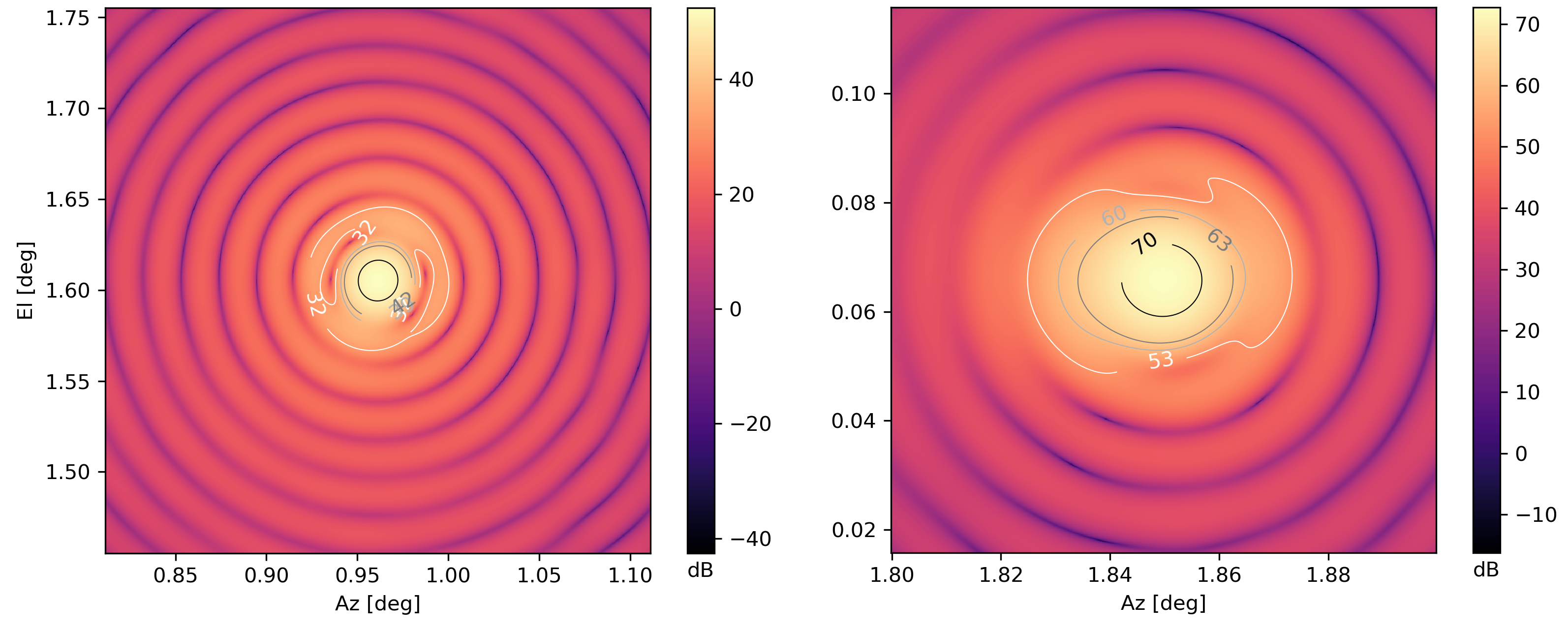}
\caption{Output of physical optics simulations showing predicted beam shape for a 150-GHz pixel (left) and a 270-GHz pixel (right) in two of the off-center optics tubes (configurations 5 and 3, respectively). The contours identify the $-3$, $-10$, $-13$, and $-20\,\rm dB$ relative to main peak (normalized to forward gain). Note the different axis scaling and color for the two panels.}
\label{fig:beams}
\end{figure}

	A key optical performance characteristic is the beam full width at half maximum (FWHM).  We calculate the FWHM by fitting an elliptical Gaussian model to the far field beam map output from GRASP\cite{GRASP} physical optics simulations (example output beam shapes are shown in Fig.~\ref{fig:beams}). The FWHM is then calculated as $\mathrm{FWHM} = \sqrt{8 \log{2} \sigma_\mathrm{max} \sigma_\mathrm{min}}$, where $\sigma_\mathrm{max}$ and $\sigma_\mathrm{min}$ are the major and minor beamwidths of the best fit elliptical Gaussian, respectively.

\begin{figure}
\begin{subfigure}{.48\textwidth}
\includegraphics[width=\textwidth]{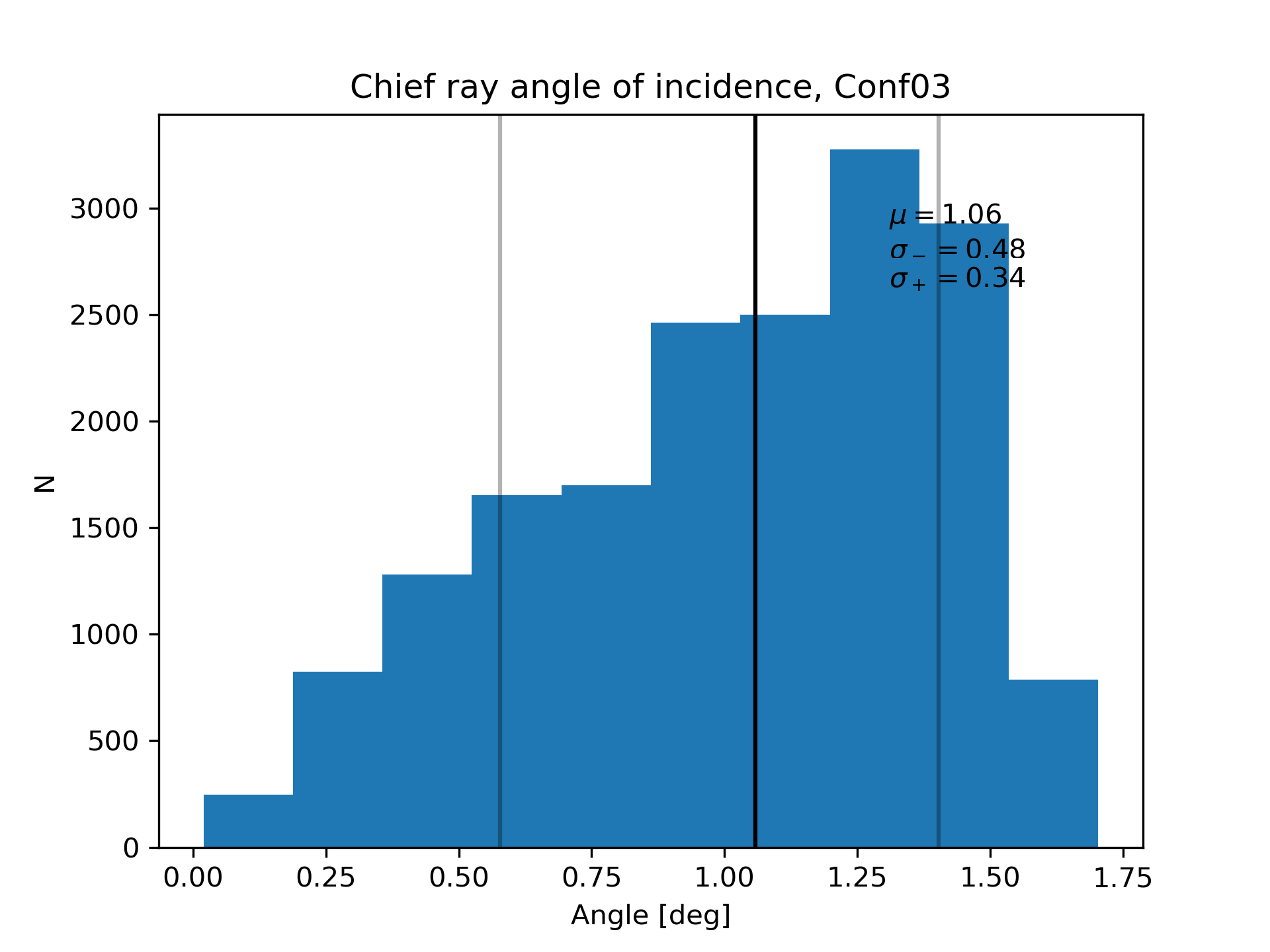}
\end{subfigure}
\begin{subfigure}{.48\textwidth}
\includegraphics[width=\textwidth]{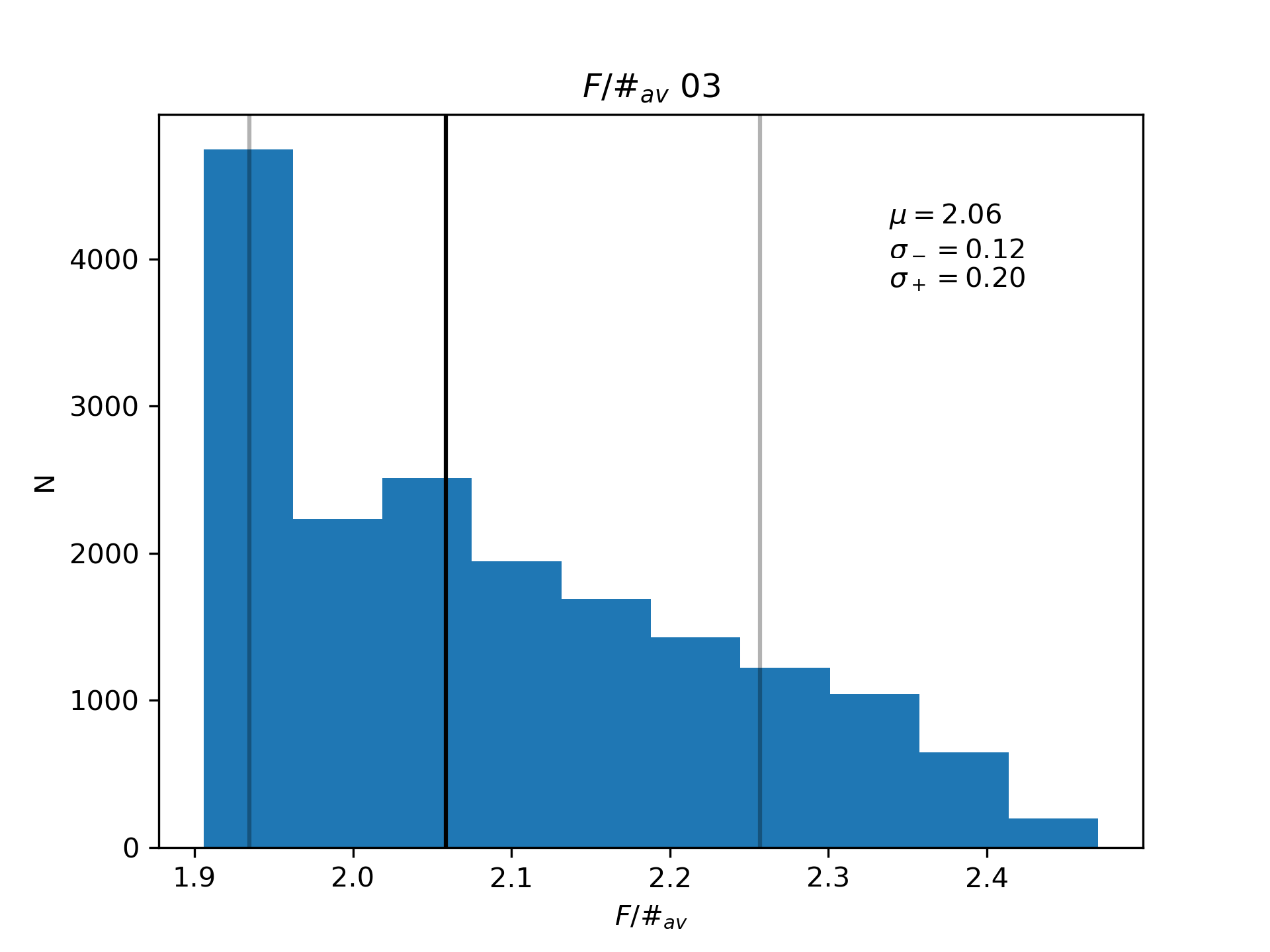}
\end{subfigure}
\caption{{\it Left:} Calculations of the chief ray angles of incidence across the focal plane of optics tube configuration 3 show that the design is well matched to feedhorn fed detector arrays with all angles below 2 degrees. {\it Right:} The variation in $f/\#$ across the focal plane of optics tube configuration 3.}
\label{fig:AOIF}
\end{figure}

With feedhorn-fed detectors, non-normal angles of incidence at the focal plane or variations in the $f/\#$ of the optics can result in beam distortions that affect the FWHM and ellipticity. The reimaging optics designs presented here and in Dicker et al. \cite{Dicker18} have been optimized for telecentricity and to minimize variations in $f/\#$. These quantities have been analyzed and compared for every optics tube configuration in over a dozen different optics designs. Examples of these analyses are shown in Figure~\ref{fig:AOIF}.

\subsection{Beam ellipticity}
	\label{sec:beamellipticity}
Non-idealities caused by the departure from beam circularity are often the first systematics considered when studying beam-related systematics. Here we quantify beam ellipticity as\footnote{A variety of definitions for beam ellipticity are used in the literature. In SO we will use this one.}
\begin{equation}
e = \frac{\sigma_\mathrm{max} - \sigma_\mathrm{min}}{\sigma_\mathrm{max} + \sigma_\mathrm{min}}.
\end{equation}
 Generally speaking, beam ellipticity can be attributed to non-uniform illumination of system apertures and non-ideal image quality (Strehl ratio); as a result, ellipticity tends to worsen as detectors are shifted away from the symmetry axis of the telescope. 

\begin{figure}
\centering
\includegraphics[width=0.98\textwidth]{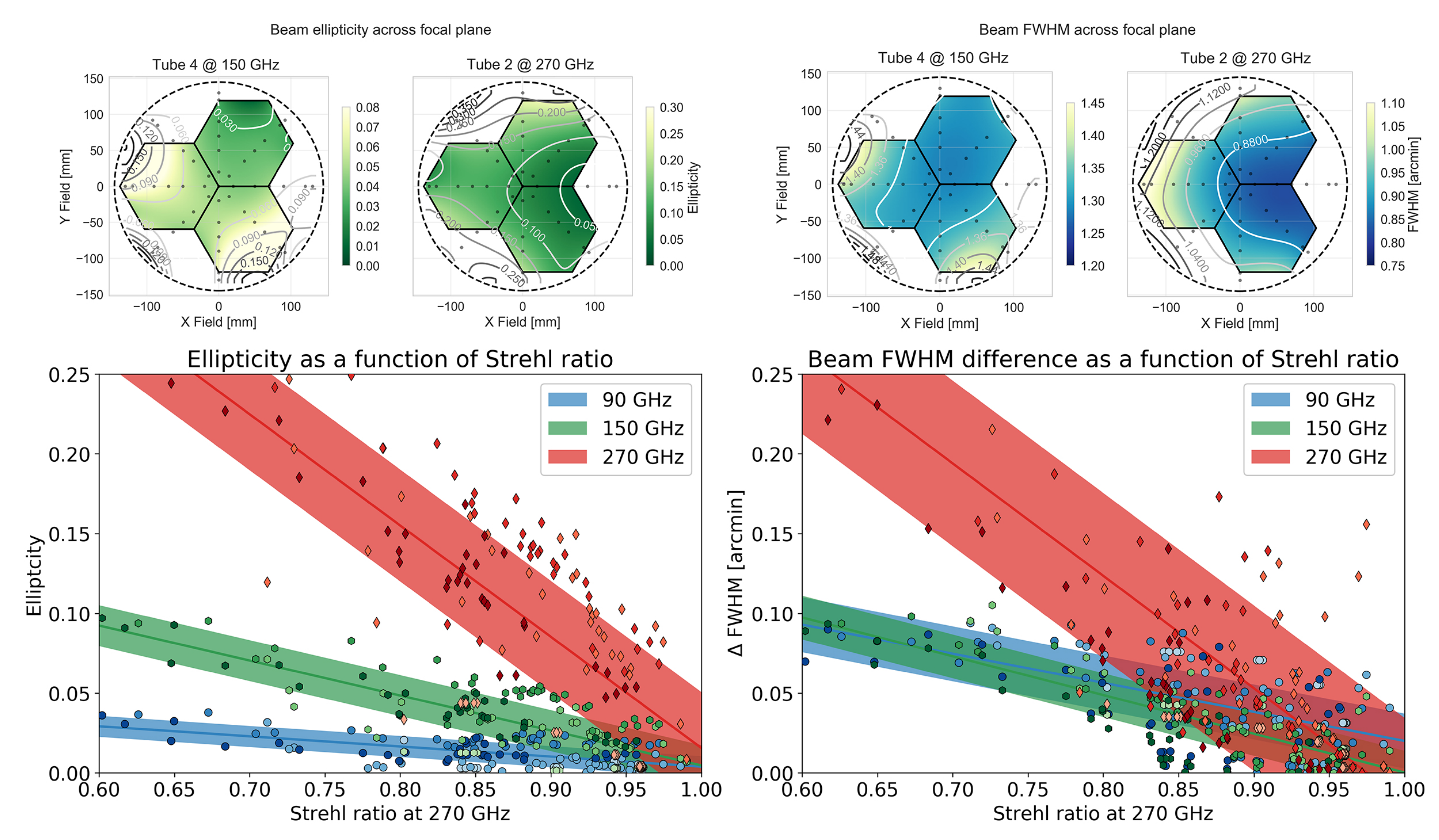}
\caption{(Top) Beam ellipticity and FWHM across focal plane for tubes 2 and 4 at 270, 150 GHz. (Bottom left) Beam ellipticity at 90, 150, and 270~GHz shown as a function of~270 GHz Strehl ratio. (Bottom right) FWHM variation at 90, 150, 270~GHz as a function of~270 GHz Strehl ratio. Four optics tubes in the SO large aperture receiver were simulated. Different shades of red, green, and blue represent different optics tubes. As Strehl ratio increases, the ellipticity and $\Delta$ FWHM decreases. Higher frequencies have a sharper dependency with frequency. Colored regions capture the best-fit line through the data points and the average per-bin standard deviation for the three frequencies (with seven bins across over the Strehl range 0.6-1.0). }
\label{fig:ecorr}
\end{figure}

While characterizing the beam response of the proposed design for the SO LAT, we see a general correlation between beam ellipticity and Strehl ratio. Figure \ref{fig:ecorr} shows the predicted beam ellipticity and FWHM of three frequency bands (90, 150, and 270~GHz) as a function of Strehl ratio at 270~GHz. As expected, beam ellipticity rises more quickly with Strehl ratio at higher frequencies. Physical optics simulations generated by GRASP were used to calculate beam ellipticities and FWHM for approximately 50 field points in four optics tubes on the LATR  (example fields are shown in Figure~\ref{fig:beams}). The corresponding Strehl ratio was calculated using Zemax OpticStudio.

In general, an asymmetric beam leads to a signal amplitude that depends on scanning angle. If not properly quantified and accounted for in analysis, this can bias both temperature and polarization measurements.\cite{Shimon2008} This systematic is most pronounced at the beam scale, but it can be somewhat remedied through a combination of cross-linking and half-wave plate modulation.

\section{Polarization Systematics}
\label{polSyst}

\subsection{Cross-polar response}			
	\label{sec:crosspolarresponse}

Cross-polarization is an optical systematic that represents the degree of polarization leakage between orthogonal polarization modes. The cross-polarization typically depends strongly on the optical design itself and describes how much the linear polarization is rotated by the reflective and refractive optical elements as light propagates through the system. If not accounted for, cross-polarization mixes the Q and U signals and can create spurious systematics in the CMB polarization signal of interest. 

The cross-polarization can be modeled using the Mueller matrix formalism\cite{hecht2002optics}. Physical optics simulations can be used to calculate the Mueller matrix beams of the telescope optical design, and these can be propagated to the power spectra using map-level simulations. A  study of the effect of detector polarization leakage in the CMB power spectra (in SO) is presented in Crowley et al.\cite{crowley18} 2018. The cross-polarization systematic leakage of the crossed-Dragone reflector system has been studied in great detail in the literature \cite{Tran2008}. The cross-polarization of re-imaging optics are highly dependent on the exact lens shapes, configuration, and anti-reflective (AR) coatings, so specific optics simulations are needed to evaluate the level of cross-polarization in a given system. It is also well known that feedhorns and lenslets can create similar levels of cross-polarization leakage and must be included in the optics simulations to achieve accurate estimates. Metal-mesh filters and anti-reflection coatings can also introduce a polarization effect. These optical elements are difficult to model, and effective parameters are measured in practice and incorporated to simulations to avoid running a physical optics into the detailed cell elements. 

The LAT has adopted a modified crossed-Dragone\cite{parshley_ccat-prime:_2018, Dicker18} design which is expected to similarly minimize the cross-polarization systematic. Because cross-polarization of the detector coupling technology can be a large source of this systematic, full physical optics and method of moments simulations of the entire optical chain (reflectors, AR coated re-imaging lenses, and baffling) with detector coupling technology specific beam inputs are planned for SO. 

Using physical optics simulations, the cross-polarization for the 6\,m class modified crossed-Dragone 
reflector design was estimated. The finite conductivity of the reflectors were assumed to be $2.5{\times}10^{7}$ S/m. The Mueller matrix representing optical properties of the reflector system was calculated from the physical optics simulations. Each Mueller matrix element represents a sky propagated Mueller beam of the telescope from a feed at the crossed Dragone focus. The results are summarized in Table \ref{Table_CDMueller}. Here the the FOV Y-axis is along the asymmetric axis of the reflector system. As expected the cross-polarization is minimal along the optical axis of the system and increases away from the FOV center. These cross-polarization values were found to be on similar levels to that of a standard crossed-Dragone design of similar aperture size and design.

\begin{table}
\begin{center}
\caption{\label{Table_CDMueller} The integrated Mueller beam powers in dB units for the instrumental polarization (IQ, IU) and cross-polarization (QU) normalized relative to the intensity beam. The Mueller matrix off-diagonal elements are symmetric and hence integrated Mueller beam powers for IQ=QI, IU=UI, and QU=UQ.}
\begin{tabular}{|c|c|c|c|}
\hline
FOV Location 				& Mueller IQ 	& Mueller IU 	& Mueller QU 	\\
\hline
Center 						& -35.12 dB		& -39.36 dB		& -40.50 dB 	\\
+Y Edge ($+3.9^{\circ}$)	& -35.45 dB		& -40.74 dB		& -26.65 dB		\\
-Y Edge ($-3.9^{\circ}$)	& -34.92 dB		& -40.29 dB		& -24.41 dB		\\
\hline
\end{tabular}
\end{center}
\end{table}

\subsection{Instrumental polarization}		 	\label{sec:instrumentalpolarizationrotation}

Instrumental polarization (IP) is a systematic that takes into account any temperature to polarization leakage that is caused by the telescope. 
This includes any polarized signal created through differential transmission or reflection by optical elements, along with any leakage resulting from pair-differencing orthogonal detectors. This systematic can be computed from theory and compared to measurements. Bryan et al. 2018 \cite{bryan18} discusses calibration techniques for future CMB observatories.

Optical IP on the LAT is dominated by the two mirrors\cite{cortiglioni_linear_1994, renbarger_measurements_1998} and the three lenses\cite{koopman_optical_2016}, and is larger at higher incidence angles. 
Because of this, IP is often worse for detectors at the edge of the focal plane. 
Polarization is introduced from the mirrors due to their finite conductivity, and the polarization fraction of the reflected light~\cite{barkats_cosmic_2005} is given by
\begin{equation}
\lambda(\nu) = 
\sqrt{4 \pi \epsilon_0 \nu \rho}\,(\sec{\theta} - \cos{\theta}),
\end{equation}
where $\rho$ is the conductivity of the metal and $\theta$ is the incident angle of the incoming light.

The IP from the modified Crossed-Dragone design calculated using physical optics simulations is summarized in Table \ref{Table_CDMueller}. These results do not include re-imaging optics. The majority of the IP is contained in the IQ Mueller term because the Q polarization was defined to be aligned along the asymmetric axis of the reflector system. It was found that compared to a reflector system with perfect conductivity, the finite conductivity ($2.5{\times}10^{7}$ S/m) of the reflectors increased the IQ power from $-43.5\pm 1.5$\,dB to -$35.5 \pm 0.5$\,dB. The IP due to the reflectors was found to be approximately similar in power across the FOV.

The IP of the lenses can be estimated using the polarization sensitive ray
trace in the optical design software Code V\cite{codev}, which was used for modeling the systematic
polarization rotation present in the Atacama Cosmology Telescope Polarimeter
(ACTPol) optical chain \cite{koopman_spie_2016}. Unpolarized fields are input,
propagating from the sky to the detector focal plane. Code V does not model the
finite conductivity of the reflectors, and so returns no contribution to the
polarization at the focal plane from the reflectors. In the case of ACTPol, the three lenses
contribute, in total, a maximum of 0.14\% near the edge of one of the off axis
focal planes. Similar analysis will be done on the SO optical design in the future.

\section{Pointing Uncertainties} 
\label{pointing}
	\label{sec:pointingerrors}
In any telescope, the optical pointing of the detectors must be accurately known and recovered over all observations through pointing reconstruction. The pointing uncertainty is any residual reconstruction difference in the pointing model due to statistical fluctuations and complex systematic effects. Hence the pointing uncertainty consists of two components: (1) statistical pointing jitter and (2) systematic pointing uncertainty and distortion.

\subsection{Statistical Pointing Jitter}
Statistical pointing jitter is the random root mean square (RMS) pointing fluctuation in the pointing reconstruction. The random pointing jitter can be caused by mechanical and structural limitations of the telescope and  residual pointing reconstruction. Telescope mechanical limitations can include telescope mechanical motion resolution, encoder resolution, structural stability, and vibrations of optical elements. Pointing jitter from analysis includes statistical pointing model fit errors due to a limited number of pointing data points in analysis.

In the case that the pointing uncertainty is random, the pointing jitter is modeled as a smearing or broadening of the instrument beam. This is equivalent to a reduction of the telescope sky resolution. This can be modeled in terms of the effect on $B_{\ell}$, the window function representing the effects of beam resolution and smoothing in the multi-pole domain. The pointing jitter is effectively a degradation of the window function and is given by
\begin{equation}
B_{\ell}^{\rm eff} = B_{\ell} e^{\frac{-\ell(\ell+1)}{2} \sigma^{2}_{p,\rm RMS}}.
\end{equation}

\begin{figure}
\centering
\includegraphics[width=.48\textwidth]{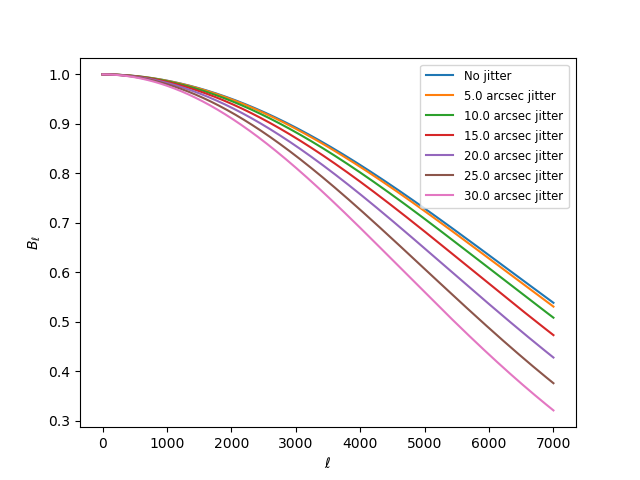}
\includegraphics[width=.48\textwidth]{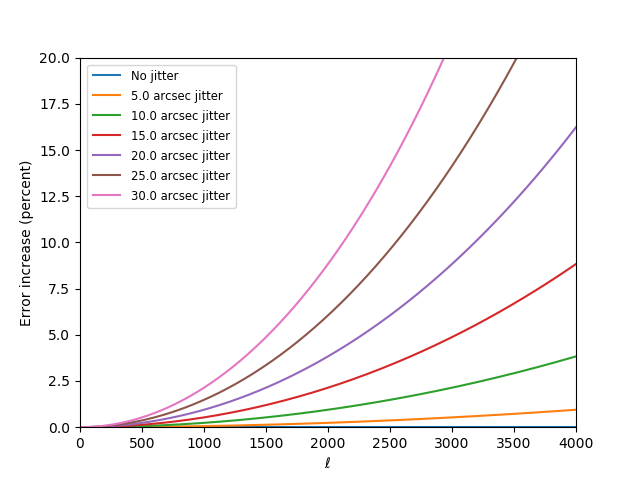}
\caption{The analytical $B_{\ell}$ uncertainty as function of pointing jitter for LAT (left), and the increase in power spectra uncertainty due to the blurring of the beam (right). The random pointing jitter increases the spectral uncertainty with increasing $\ell$. For typical 10 arcsec pointing jitter RMS achieved by past large aperture experiments, the error increases to $>1\%$ above $\ell>2000$.} 
\label{fig:b_ell_error}
\end{figure}

Here $\sigma^{2}_{p,RMS}$ is the statistical RMS pointing reconstruction uncertainty. This statistical pointing jitter results in increased statistical uncertainty in the power spectra $C_{\ell}$ at high $\ell$ as a function of the original beam size and RMS value. The theoretical uncertainty increase as a function of $\ell$ for the LAT is shown in Fig.~\ref{fig:b_ell_error}. Typically large aperture experiments are designed to achieve $\leq10$ arcsec levels of pointing RMS jitter by designing in structural stability of optical elements as well as encoder readout stability. For example this would equate to $<4\%$ reduction in sensitivity at $\ell\leq4000$. With careful pointing reconstruction analysis using point sources and/or star cameras, this jitter can potentially be reduced further.

\subsection{Systematic Pointing Uncertainty and Distortion}
Systematic pointing uncertainty and distortion are the residual uncertainties in the pointing reconstruction due to external parameters creating complex flexure in the telescope structure which leads to misalignments and deformations of the instrument optics. These external parameters represent any change in the telescope environment such as ambient temperature, solar irradiance, and structural deterioration. External parameters that correlate with the sun and weather will change on hourly time scales as well as throughout the year. Structural deterioration will occur on much longer time scales of months to years and this pointing change may be negligible within a single observation season.

Here we have made a distinction where the systematic pointing error refers to misalignments between optical elements and pointing distortion refers to deformations of the optical elements. Both effects depend highly on the telescope structural design and are difficult to model exactly.\cite{parshley_ccat-prime:_2018,parshley_optical_2018,Dicker18} Misalignments can be approximated using geometric optics tolerance models. Deformations are comparatively more difficult to model because deformations can change the F-number of each optical element due to changes in the conical surface shapes which are typically difficult to measure frequently.

As done in \textsc{Polarbear} \cite{polarbear17}, the pointing reconstruction model can be expanded to include systematic effects due to external parameters. Two additional terms were added to model the structural flexure from varying ambient temperature. Another two terms were used to model differential solar heating across different parts of the telescope dependent on sun position and telescope orientation.

The systematic pointing error and distortion cannot be modeled simply by a smearing of the instrument beam, and will generally create asymmetric non-Gaussian effective instrument beam shapes. Detailed measurements (such as photogrammetry) of metrology and surface shapes across different environmental conditions and physical optics simulations will be needed to model such systematic pointing model effects. The error in the power spectra can be estimated through simulations comparing pointing models with different sets of systematics parameters as done in \textsc{Polarbear} \cite{polarbear17}. 

Modeling complex environmental effects is difficult, but these errors can typically be minimized through various means. Added mechanical implementations such as structural reinforcements, baffling, and insulation are effective. Environmental regulation of key components (i.e. receiver mount) can also reduce the systematics. Active monitoring of structural temperature gradients and vibrations can aid in modeling these pointing systematics as well. With necessary precautions these pointing systematics can be minimized effectively.

\section{Sidelobe pickup}
	\label{sec:sidelobepickup}
The angular response of an imaging optical system is described by its power pattern which consists of a main lobe, where the angular response of the system is maximum, and sidelobes. Sidelobes result from non-ideal interactions in the system and can be loosely classified into near sidelobes and far sidelobes.

Near sidelobes are close ($\sim \frac{\lambda}{D}$ to 1 field of view (FOV)) to the main beam. Their origin can be diffractive, can be the result of multiple reflections in the optical path or it can result from scattering in an optical component. When  reflections occur inside of the refractive camera, they can result in ghosting, where a secondary image forms from the primary image's reflections. Polarized near sidelobes have been observed in ACT that follow a square cross pattern \cite{louis_atacama_2017}. These are consistent with diffraction off a metal mesh  filter at the Lyot stop which will be avoided in SO. 

Far sidelobes are located at large angles from the main beam (from 1 FOV to 180 degrees). They are typically more difficult to model and characterize than near sidelobes and are generated by a variety of physical mechanisms like reflections, diffraction, scattering or a combination of them. Diffraction can occur if the paneled surface of the primary and secondary mirrors have gaps with a size comparable to a wavelength. This effect was studied in ACT \cite{fluxa_rojas_far_2016}. Scattering at the camera can increase the sensitivity of the camera to stray light, generating large angle sidelobes\cite{gallardo18}.

The design of high-sensitivity transition-edge sensor (TES) coupled experiments requires both low sidelobes and low spillover. Large angle sidelobes are undesirable as they inject faint signals of objects far away from the main beam, while spillover degrades detector performance due to increased photon noise. Spilled or thermal radiation can have a potentially severe impact on detector performance and can be so severe that the optical power surpasses detector saturation power. Introducing reflective baffles that guide rays (in the time-reverse sense) from the receiver camera out to the sky instead of terminating them at an ambient temperature surface trades thermal light for sidelobes.

In this study, we model near sidelobes from ghosting and far sidelobes from multiple reflections on the telescope structure.

\begin{figure}
\centering
\begin{subfigure}{0.48\textwidth}
\includegraphics[width=\textwidth]{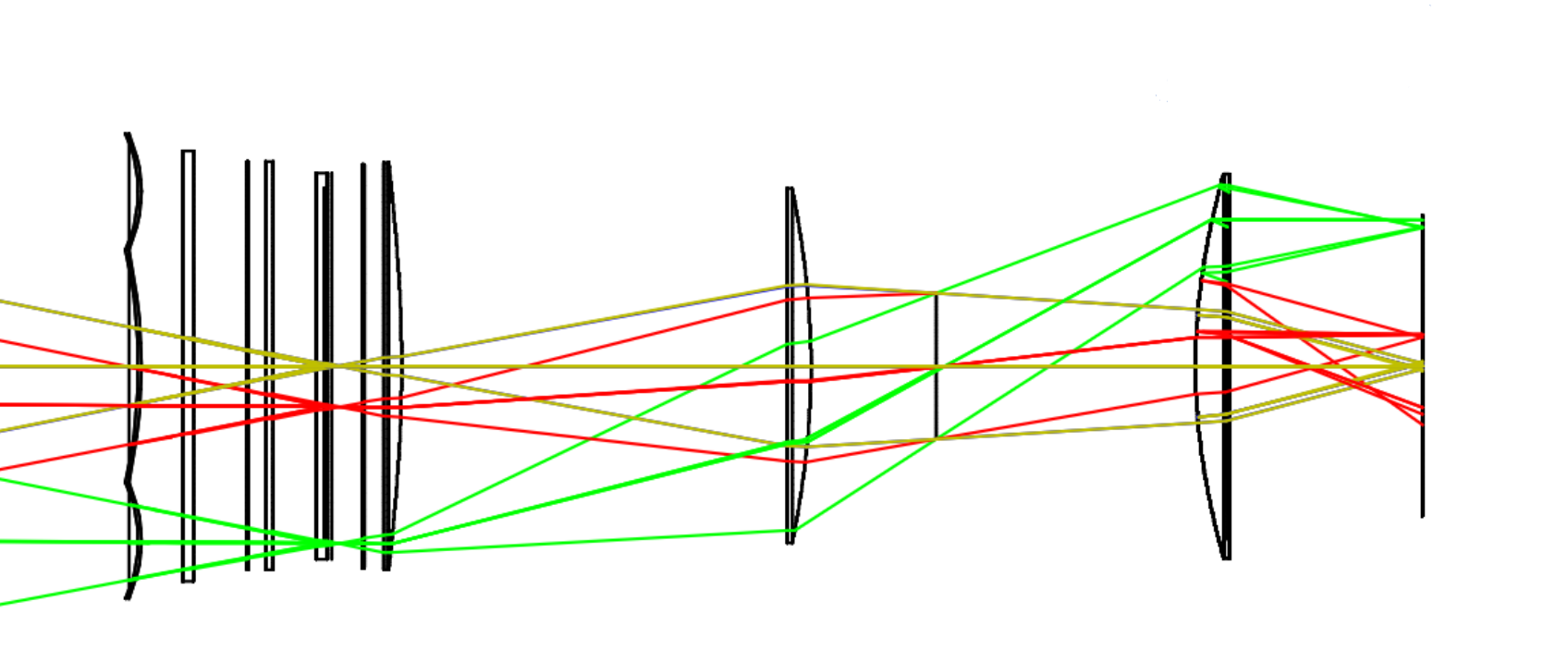}
\end{subfigure}
\begin{subfigure}{0.48\textwidth}
\includegraphics[width=\textwidth]{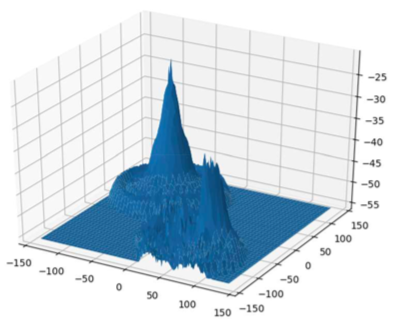}
\end{subfigure}

\caption{Left: Non-sequential ray trace showing an example of a ghost image. Rays from a field close to the center (red) reflect off the focal plane, and get reflected on one of the lens surfaces to form a spurius image like the one shown on the right. Rays for other field positions (green and gold) are also shown. Right: Ghosting image intensity for an off-axis pixel in the LAT design. This is a pessimistic estimate driven by the 1\% reflectivity of the AR coatings in the model and the perfect reflectivity of the focal plane. In practice AR coatings have been demonstrated with lower reflectivity\cite{datta_large-aperture_2013}, which will decrease the amplitude of the ghosted signal. Vertical axis is shown in dB from the main beam. }
\label{fig:ghosting}
\end{figure}

\subsection{Near sidelobes from ghosting}

Ghost images are generated by small reflections off optical elements and are a source of near sidelobes. Ghosting is simulated using a ray-tracing software (Zemax OpticStudio) in the non-sequential mode. Here, we model the optical chain (two-mirror system, three-lens camera with lyot stop, filters and window) allowing for reflections inside the optics tube. Reflections in the lens interfaces are allowed at the 1\% level and the focal plane is modeled with a reflectance of 100\%. Plastic windows and filters are modeled with 0.5\% reflectance per interface. The inner boundary of the optics tube is modeled as a perfect absorber such that rays that escape the tube on the sides do not reflect back. The diffraction PSF  from Huygens theory of diffraction is superimposed to the ghosting pattern to show the scale of the ghosting. Figure \ref{fig:ghosting} shows that the ghost image for the center field is diffuse and has an intensity ranging from $-25\,\rm dB$ (near the main beam) to $-50\,\rm dB$ (shown as a halo around it) below the main beam. This is a pessimistic value (due to the values chosen for the reflectivity of the AR coatings and the reflectivity of the focal plane) intended to be interpreted as an upper bound; the actual ghosting intensity will depend on the reflectivities of the refractive system.

\begin{figure}

\begin{subfigure}{.48\textwidth}
\includegraphics[width=\textwidth]{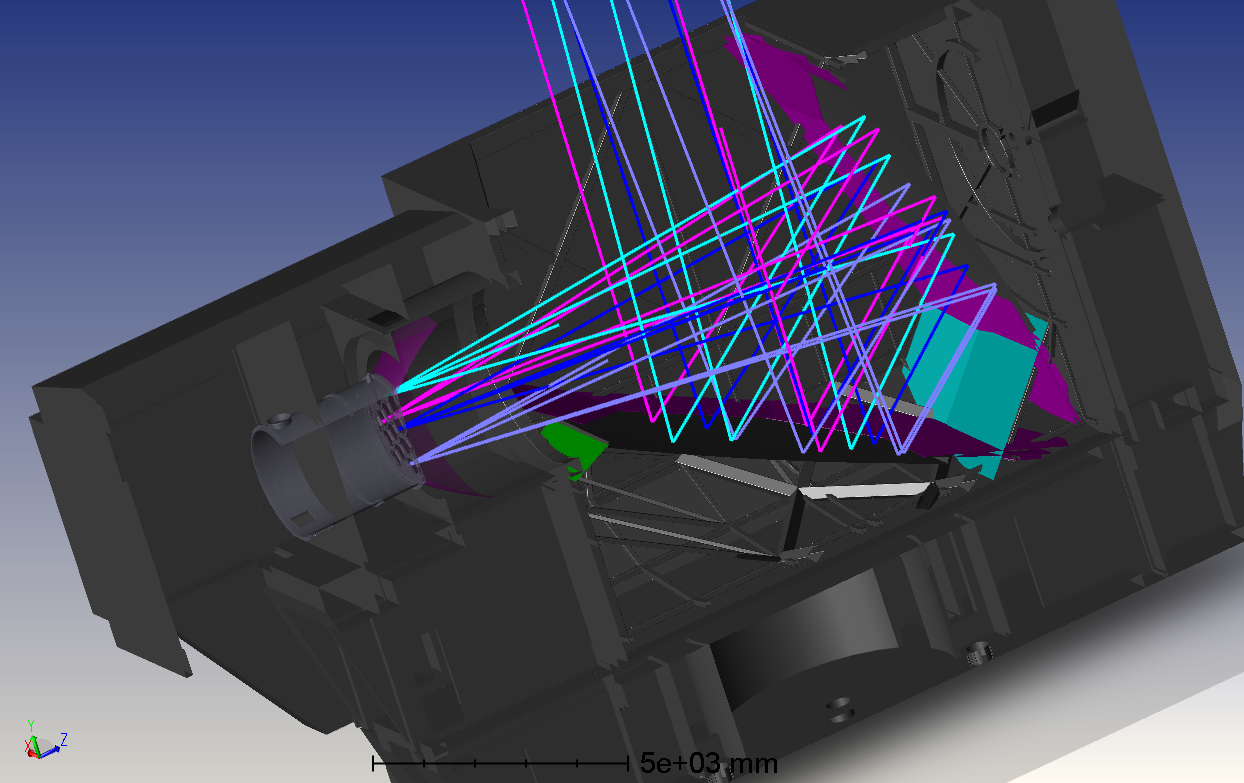}
\end{subfigure}
\begin{subfigure}{.48\textwidth}
\includegraphics[width=\textwidth]{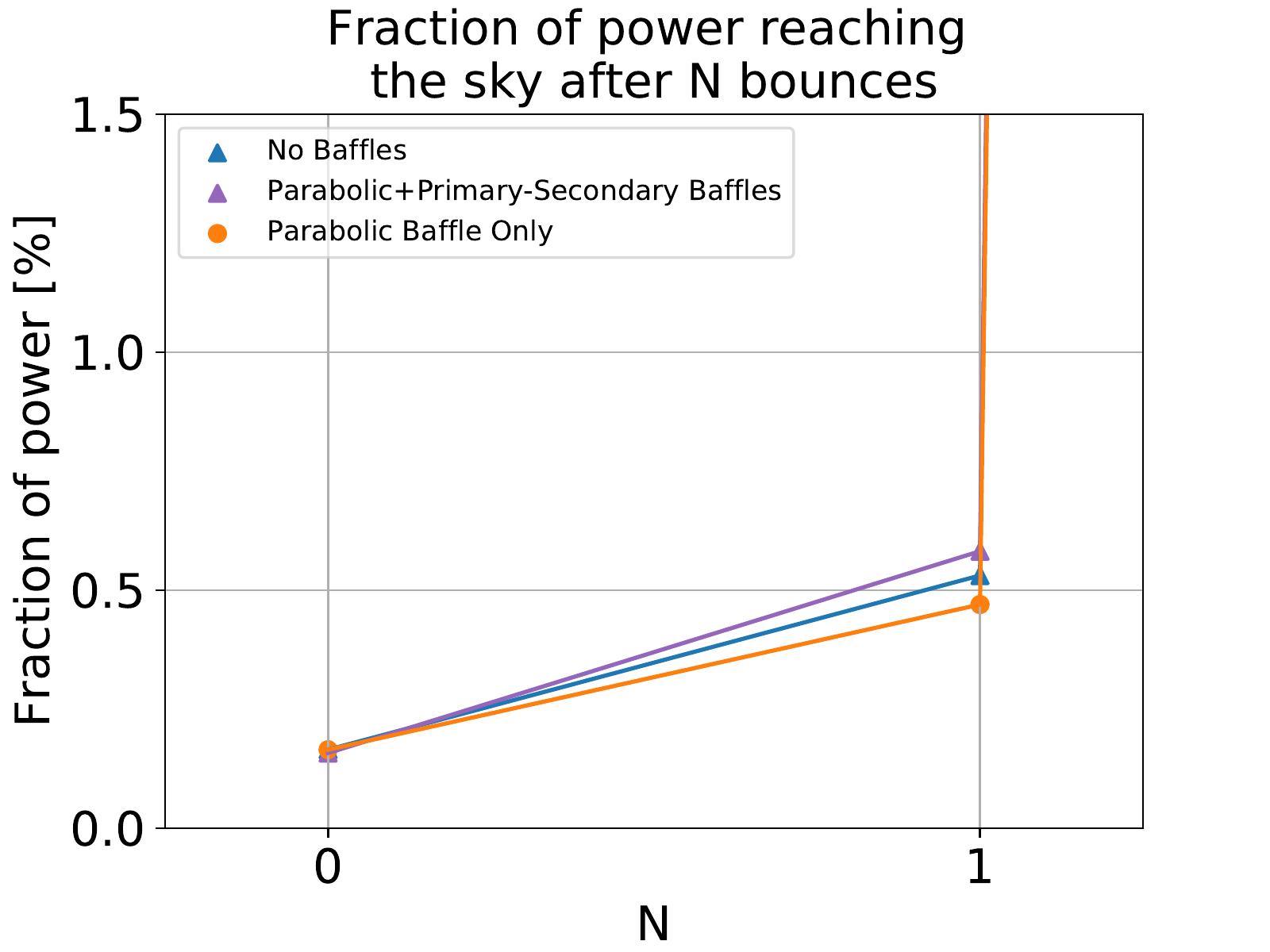}
\end{subfigure}

\begin{subfigure}{.48\textwidth}
\includegraphics[width=\textwidth]{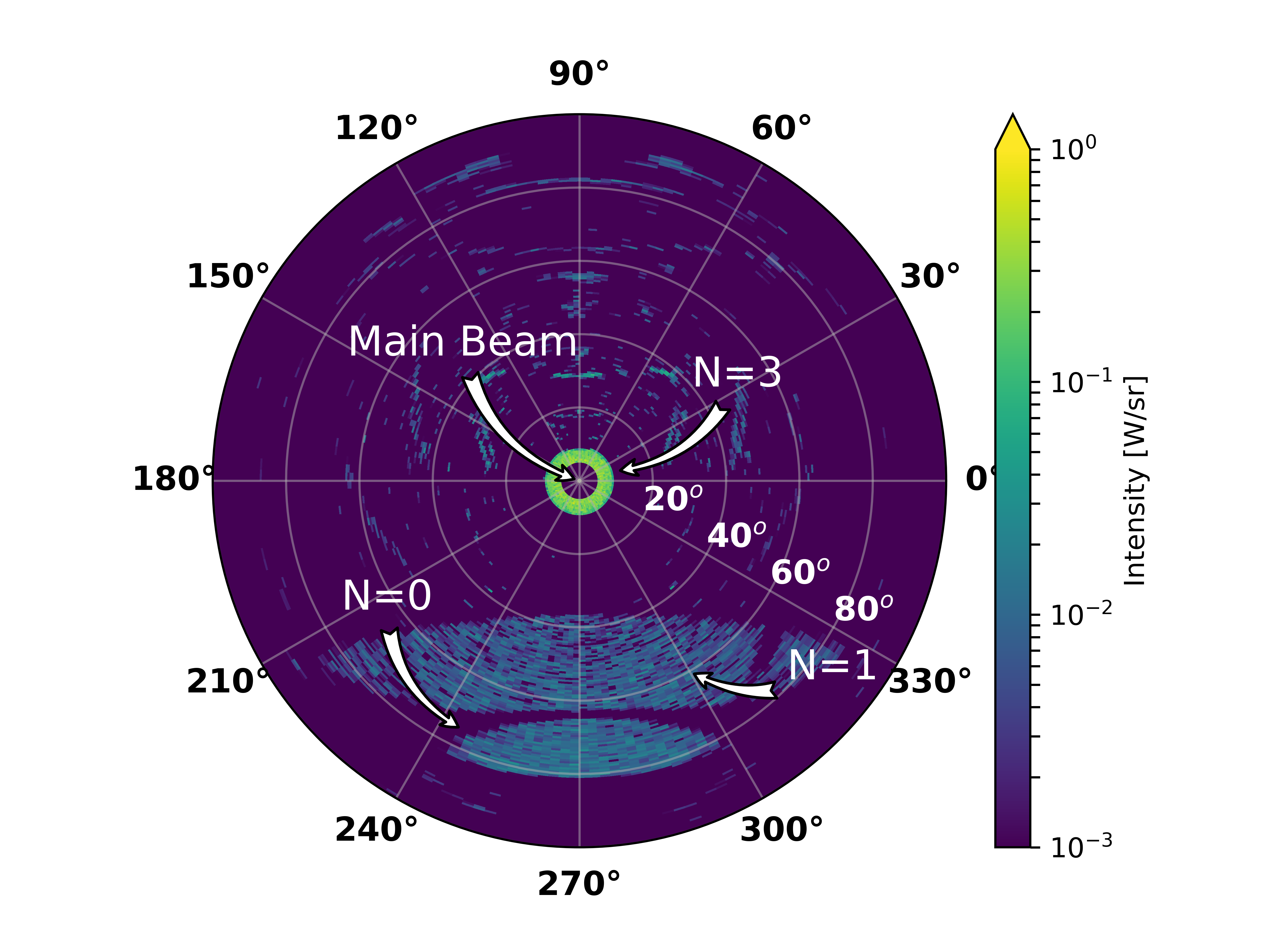}
\end{subfigure}
\begin{subfigure}{.48\textwidth}
\includegraphics[width=\textwidth]{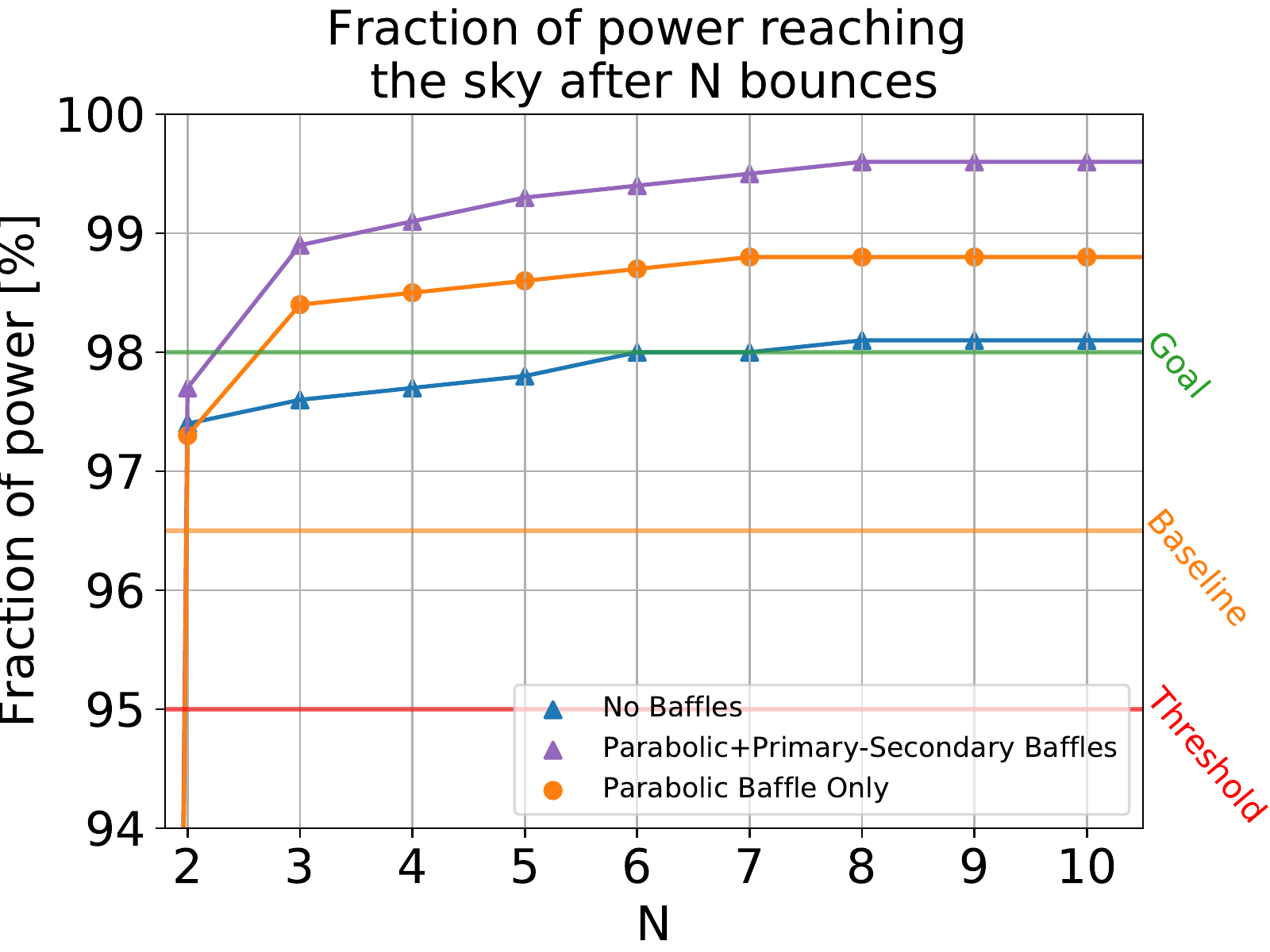}
\end{subfigure}

\caption{Non-sequential ray trace model for the large aperture telescope. 
Upper left shows configuration c (parabolic baffle with a large secondary and primary guard ring baffles) and the position of 4 fields being launched from the focal plane of the telescope. The model contains primary and secondary mirrors, a parabolic baffle around the camera aperture (pink),  guard rings around the primary and secondary mirrors (also pink), a flat reflector covering the space on the back of the primary mirror (green) and a three-segment reflector placed between the primary and secondary reflectors (cyan). This configuration is meant to serve as an upper limit, more reduced baffles will be implemented in SO. Upper right (lower right) shows the fractional power that reaches the sky from the center field of the center camera versus number of bounces for a perfectly reflective telescope solid model for N in the interval 0-1  (2-10) bounces. Lower left shows the power density at the sky for N=3 bounces for configuration b (parabolic baffle alone). Power density at the sky is normalized to have an injected power equal to $1\,\rm W$. In this configuration, one sidelobe is expected between 60 and 80 degrees, corresponding to the direct line of sight from the receiver camera to the sky. The parabolic baffle forms a ring around the main beam at N=3, which contains most of the difference in power between the parabolic baffle and the no-baffle model.}
\label{fig:ray_trace}
\end{figure}

\subsection{Far sidelobes}

The far sidelobe response of the system is simulated using the hybrid approach studied in ACTPol \cite{gallardo18}. Here the beam of the camera receiver is modeled using a phenomenological fit obtained from measurements in the field. This approach is preferred over a full physical optics model, as it is faster to run and captures effects that are hard to model from diffraction alone. Other effects that are difficult to model in a physical optics simulation include scattering off optical elements and diffuse reflections off the inner walls of the camera optics tube.

\begin{table}[]
\centering
\begin{tabular}{l|cccc}
Parameter & $\alpha [\rm deg]$ & $\beta\rm [deg]$ & A [-]     & $\theta_o \rm[deg]$ \\
\hline
Value & 12.7     & 38      & 1.9$\times10^{-3}$ & 12        
\end{tabular}
\caption{Parameters used in the beam fit model. The intensity of the camera beam is described by Eq. \ref{eq:NickModel}, where $\alpha$ and $\beta$ control the decay of the Gaussian and exponential functions as a function of angle. A is the relative intensity of the exponential tails relative to the center beam and $\theta_o$ is the transition angle, given by the Lyot stop diameter and the optics of the camera. \cite{gallardo18}}
\label{tab:beamParams}
\end{table}

The camera beam is modeled using a Gaussian beam with exponential tails that represent the level of scattering that the optical elements present to incoming light. In this model the response of the three-lens system to incoming light at an angular distance $\theta$ is given by\cite{gallardo18} 
\begin{equation}
\label{eq:NickModel}
        f(\theta)=
        \left\{ \begin{array}{ll}
            \exp{-(\theta/\alpha)^2} &\theta \leq \theta_o \\
            A \exp{-(\theta/\beta)} &\theta > \theta_o 
        \end{array} \right.
\end{equation}
where $\theta_o$ is the secondary mirror illumination angle and is given by the F-number of the primary-secondary system. $\alpha$ and $\beta$ are the angular parameters setting the width of the Gaussian profile and the speed of decay of the tails of the distribution. Parameter $A$ is a scale factor that sets the overall level of the exponential tails relative to the central Gaussian profile. Values in equation \ref{eq:NickModel} are taken from measurements performed for AdvACT \cite{gallardo18}. Table \ref{tab:beamParams} shows the numerical values used in this simulation. The AdvACT camera has enough similarities to the proposed SO camera at millimeter wavelengths, that the use of the ACT beam model is justified until SO instrument beam measurements are available. While the recently measured sidelobe levels appear promising (3~\%), additional studies are in progress to improve the spilled light at the receiver and effectively reduce the parameter $A$ in the equivalent beam model presented here. 

The beam model is simulated in a ray-tracing software (Zemax OpticStudio\cite{zemax_llc_zemax_2016}), where the intensity determines the probability of a ray being realized in one direction. We use a CAD 3D model of the mechanical structure of the telescope and the reflective part of the optics to predict how much light gets from the camera window to the sky (and in what direction). In practice, this simulation launches rays from one position on the camera window surface with the given beam model described in equation \ref{eq:NickModel} and counts how many rays get to the sky after a finite number of bounces inside the telescope structure. This method allows for rapid evaluation of different prototypes.

We have compared three configurations: one with no baffling (a), one with a parabolic baffle around the receiver window (b) and another configuration with a parabolic baffle, a guard ring around primary and secondary reflectors plus a three segment baffle between primary and secondary (c). Configuration c is merely conceptual as it extends physically as much as possible inside the elevation structure and it serves here as an upper limit of how well this design could perform. A reduced version of the primary-secondary baffle will be implemented in SO. Figure \ref{fig:ray_trace} (upper left) shows one model (configuration c) with a square aperture at the entrance of the system, a parabolic baffle around the receiver aperture (shown in pink), a reflective baffle (made out of flat segments) around the primary and secondary mirrors (also pink), a three segment flat panel baffle on the lower part of the primary/secondary corner (cyan) and a flat baffle covering the back of the primary mirror (green). 

The figure of merit used is the fraction of power reaching the sky after N bounces for each model configuration. The mapping speed of the experiment is a steep function of the spillover fraction as discussed in Hill et al. 2018\cite{hill_bolocalc:_2018} (section 4.4). Based in these mapping speed studies, we have established that the SO sensitivity goal corresponds to a 98\% spillover fraction while the threshold for operation corresponds to 95\%, we define a baseline level at 96.5\%. Early ACT beam measurements suggested that SO might have trouble meeting the goal; however, our most recent and precise measurements suggest that our spillover goal is within reach. Figure \ref{fig:ray_trace}  (lower right) shows this and the effect of the three studied configurations. The nominal (a) configuration with no baffles reaches 97.4\% of the power at the sky after two reflections. As much as 1\% in spillover can be gained at the third reflection if a parabolic baffle alone is used. In this model the upper limit (configuration c) shows less than a one percent improvement from the parabolic baffle alone configuration (b).

The angular distribution of the scattered light at the sky can also be extracted from this model. Figure \ref{fig:ray_trace} (lower left) shows that around the main beam a ring at around 10 degrees appears as a consequence of the parabolic baffle, also a spillover sidelobe forms at large angles (N=0) between 70 and 80 degrees. The first bounce feature is extended and it appears between 40 and 70 degrees from the main beam. This predicted sidelobe pattern will be included in the design of our observation strategy.\cite{stevens_designs_2018}

\section{Conclusion and future work}
	\label{sec:conclusion}
SO will measure the CMB with unprecedented sensitivity. Such an increase of sensitivity puts greater challenges on how we understand and control systematic effects. In this paper, we present some of the computational tools we are using to model and incorporate these high order effects into the design of the instrument. These tools have  progressed substantially in recent decades, allowing for simulations of complex systems with a great level of accuracy and enabling rapid iterations of prototype designs and implementations. These tools will not only benefit the design of the instrument, but they are also being used to determine calibration requirements. The tools and methods developed for SO and presented here can be extended to the design of future experiments such as CMB-S4.

\acknowledgments 
This work was supported in part by a grant from the Simons Foundation (Award \#457687, B.K.)

\nocite{*} 
\bibliography{report}

\begin{thebibliography}{10}

\bibitem{ahmed_bicep3:_2014}
Ahmed, Z., Amiri, M., Benton, S.~J., Bock, J.~J., Bowens-Rubin, R., Buder, I.,
  Bullock, E., Connors, J., Filippini, J.~P., Grayson, J.~A., Halpern, M.,
  Hilton, G.~C., Hristov, V.~V., Hui, H., Irwin, K.~D., Kang, J., Karkare,
  K.~S., Karpel, E., Kovac, J.~M., Kuo, C.~L., Netterfield, C.~B., Nguyen,
  H.~T., O'Brient, R., Ogburn, R.~W., Pryke, C., Reintsema, C.~D., Richter, S.,
  Thompson, K.~L., Turner, A.~D., Vieregg, A.~G., Wu, W. L.~K., and Yoon,
  K.~W., ``{BICEP}3: a 95ghz refracting telescope for degree-scale {CMB}
  polarization,'' in [{\em Millimeter, {Submillimeter}, and {Far}-{Infrared}
  {Detectors} and {Instrumentation} for {Astronomy}
  {VII}}{\nolinebreak\hspace{0.1em}]},   {\bf 9153},  91531N, International
  Society for Optics and Photonics (Aug. 2014).

\bibitem{thornton_atacama_2016}
Thornton, R.~J., Ade, P. A.~R., Aiola, S., Angile, F.~E., Amiri, M., Beall,
  J.~A., Becker, D.~T., Cho, H.-M., Choi, S.~K., Corlies, P., Coughlin, K.~P.,
  Datta, R., Devlin, M.~J., Dicker, S.~R., Dunner, R., Fowler, J.~W., Fox,
  A.~E., Gallardo, P.~A., Gao, J., Grace, E., Halpern, M., Hasselfield, M.,
  Henderson, S.~W., Hilton, G.~C., Hincks, A.~D., Ho, S.~P., Hubmayr, J.,
  Irwin, K.~D., Klein, J., Koopman, B., Li, D., Louis, T., Lungu, M., Maurin,
  L., McMahon, J., Munson, C.~D., Naess, S., Nati, F., Newburgh, L., Nibarger,
  J., Niemack, M.~D., Niraula, P., Nolta, M.~R., Page, L.~A., Pappas, C.~G.,
  Schillaci, A., Schmitt, B.~L., Sehgal, N., Sievers, J.~L., Simon, S.~M.,
  Staggs, S.~T., Tucker, C., Uehara, M., van Lanen, J., Ward, J.~T., and
  Wollack, E.~J., ``The {Atacama} {Cosmology} {Telescope}: {The}
  polarization-sensitive {ACTPol} instrument,'' {\em arXiv:1605.06569
  [astro-ph]}  (May 2016).
\newblock arXiv: 1605.06569.

\bibitem{benson2014spt}
Benson, B.~A., Ade, P. a.~R., Ahmed, Z., Allen, S.~W., Arnold, K., Austermann,
  J.~E., Bender, A.~N., Bleem, L.~E., Carlstrom, J.~E., Chang, C.~L., Cho,
  H.~M., Cliche, J.~F., Crawford, T.~M., Cukierman, A., Haan, T.~d., Dobbs,
  M.~A., Dutcher, D., Everett, W., Gilbert, A., Halverson, N.~W., Hanson, D.,
  Harrington, N.~L., Hattori, K., Henning, J.~W., Hilton, G.~C., Holder, G.~P.,
  Holzapfel, W.~L., Irwin, K.~D., Keisler, R., Knox, L., Kubik, D., Kuo, C.~L.,
  Lee, A.~T., Leitch, E.~M., Li, D., McDonald, M., Meyer, S.~S., Montgomery,
  J., Myers, M., Natoli, T., Nguyen, H., Novosad, V., Padin, S., Pan, Z.,
  Pearson, J., Reichardt, C., Ruhl, J.~E., Saliwanchik, B.~R., Simard, G.,
  Smecher, G., Sayre, J.~T., Shirokoff, E., Stark, A.~A., Story, K., Suzuki,
  A., Thompson, K.~L., Tucker, C., Vanderlinde, K., Vieira, J.~D., Vikhlinin,
  A., Wang, G., Yefremenko, V., and Yoon, K.~W., ``{SPT}-3g: a next-generation
  cosmic microwave background polarization experiment on the {South} {Pole}
  telescope,'' in [{\em Millimeter, {Submillimeter}, and {Far}-{Infrared}
  {Detectors} and {Instrumentation} for {Astronomy}
  {VII}}{\nolinebreak\hspace{0.1em}]},   {\bf 9153},  91531P, International
  Society for Optics and Photonics (July 2014).

\bibitem{suzuki2016polarbear}
Suzuki, A., Ade, P., Akiba, Y., Aleman, C., Arnold, K., Baccigalupi, C., Barch,
  B., Barron, D., Bender, A., Boettger, D., Borrill, J., Chapman, S., Chinone,
  Y., Cukierman, A., Dobbs, M., Ducout, A., Dunner, R., Elleflot, T., Errard,
  J., Fabbian, G., Feeney, S., Feng, C., Fujino, T., Fuller, G., Gilbert, A.,
  Goeckner-Wald, N., Groh, J., Haan, T.~D., Hall, G., Halverson, N., Hamada,
  T., Hasegawa, M., Hattori, K., Hazumi, M., Hill, C., Holzapfel, W., Hori, Y.,
  Howe, L., Inoue, Y., Irie, F., Jaehnig, G., Jaffe, A., Jeong, O., Katayama,
  N., Kaufman, J., Kazemzadeh, K., Keating, B., Kermish, Z., Keskitalo, R.,
  Kisner, T., Kusaka, A., Jeune, M.~L., Lee, A., Leon, D., Linder, E., Lowry,
  L., Matsuda, F., Matsumura, T., Miller, N., Mizukami, K., Montgomery, J.,
  Navaroli, M., Nishino, H., Peloton, J., Poletti, D., Puglisi, G., Rebeiz, G.,
  Raum, C., Reichardt, C., Richards, P., Ross, C., Rotermund, K., Segawa, Y.,
  Sherwin, B., Shirley, I., Siritanasak, P., Stebor, N., Stompor, R., Suzuki,
  J., Tajima, O., Takada, S., Takakura, S., Takatori, S., Tikhomirov, A.,
  Tomaru, T., Westbrook, B., Whitehorn, N., Yamashita, T., Zahn, A., and Zahn,
  O., ``The {Polarbear}-2 and the {Simons} {Array} {Experiments},'' {\em
  Journal of Low Temperature Physics}~{\bf 184},  805--810 (Aug. 2016).

\bibitem{planck_collaboration_planck_2018}
Collaboration, P., Akrami, Y., Arroja, F., Ashdown, M., Aumont, J.,
  Baccigalupi, C., Ballardini, M., Banday, A.~J., Barreiro, R.~B., Bartolo, N.,
  Basak, S., Benabed, K., Bernard, J.-P., Bersanelli, M., Bielewicz, P., Bock,
  J.~J., Bond, J.~R., Borrill, J., Bouchet, F.~R., Boulanger, F., Bucher, M.,
  Burigana, C., Butler, R.~C., Calabrese, E., Cardoso, J.-F., Carron, J.,
  Challinor, A., Chiang, H.~C., Colombo, L. P.~L., Combet, C., Contreras, D.,
  Crill, B.~P., Cuttaia, F., de~Bernardis, P., de~Zotti, G., Delabrouille, J.,
  Delouis, J.-M., Di~Valentino, E., Diego, J.~M., Donzelli, S., DorÃ©, O.,
  Douspis, M., Ducout, A., Dupac, X., Dusini, S., Efstathiou, G., Elsner, F.,
  EnÃŸlin, T.~A., Eriksen, H.~K., Fantaye, Y., Fergusson, J.,
  Fernandez-Cobos, R., Finelli, F., Forastieri, F., Frailis, M., Franceschi,
  E., Frolov, A., Galeotta, S., Galli, S., Ganga, K., Gauthier, C.,
  GÃ©nova-Santos, R.~T., Gerbino, M., Ghosh, T., GonzÃ¡lez-Nuevo, J.,
  GÃ³rski, K.~M., Gratton, S., Gruppuso, A., Gudmundsson, J.~E., Hamann, J.,
  Handley, W., Hansen, F.~K., Herranz, D., Hivon, E., Hooper, D.~C., Huang, Z.,
  Jaffe, A.~H., Jones, W.~C., KeihÃ¤nen, E., Keskitalo, R., Kiiveri, K., Kim,
  J., Kisner, T.~S., Krachmalnicoff, N., Kunz, M., Kurki-Suonio, H., Lagache,
  G., Lamarre, J.-M., Lasenby, A., Lattanzi, M., Lawrence, C.~R., Jeune, M.~L.,
  Lesgourgues, J., Levrier, F., Lewis, A., Liguori, M., Lilje, P.~B., Lindholm,
  V., Lpez-Caniego, M., Lubin, P.~M., Ma, Y.-Z., MacÃ­as-PÃ©rez, J.~F.,
  Maggio, G., Maino, D., Mandolesi, N., Mangilli, A., Marcos-Caballero, A.,
  Maris, M., Martin, P.~G., MartÃ­nez-GonzÃ¡lez, E., Matarrese, S., Mauri,
  N., McEwen, J.~D., Meerburg, P.~D., Meinhold, P.~R., Melchiorri, A.,
  Mennella, A., Migliaccio, M., Mitra, S., Miville-DeschÃªnes, M.-A.,
  Molinari, D., Moneti, A., Montier, L., Morgante, G., Moss, A., MÃ¼nchmeyer,
  M., Natoli, P., NÃ¸rgaard-Nielsen, H.~U., Pagano, L., Paoletti, D.,
  Partridge, B., Patanchon, G., Peiris, H.~V., Perrotta, F., Pettorino, V.,
  Piacentini, F., Polastri, L., Polenta, G., Puget, J.-L., Rachen, J.~P.,
  Reinecke, M., Remazeilles, M., Renzi, A., Rocha, G., Rosset, C., Roudier, G.,
  RubiÃ±o-MartÃ­n, J.~A., Ruiz-Granados, B., Salvati, L., Sandri, M.,
  Savelainen, M., Scott, D., Shellard, E. P.~S., Shiraishi, M., Sirignano, C.,
  Sirri, G., Spencer, L.~D., Sunyaev, R., Suur-Uski, A.-S., Tauber, J.~A.,
  Tavagnacco, D., Tenti, M., Toffolatti, L., Tomasi, M., Trombetti, T.,
  Valiviita, J., Van~Tent, B., Vielva, P., Villa, F., Vittorio, N., Wandelt,
  B.~D., Wehus, I.~K., White, S. D.~M., Zacchei, A., Zibin, J.~P., and Zonca,
  A., ``Planck 2018 results. {X}. {Constraints} on inflation,'' {\em
  arXiv:1807.06211 [astro-ph]}  (July 2018).
\newblock arXiv: 1807.06211.

\bibitem{hasselfield_atacama_2013}
Hasselfield, M., Hilton, M., Marriage, T.~A., Addison, G.~E., Barrientos,
  L.~F., {Nicholas Battaglia}, Battistelli, E.~S., Bond, J.~R., Crichton, D.,
  Das, S., Devlin, M.~J., Dicker, S.~R., Dunkley, J., DÃ¼nner, R., Fowler,
  J.~W., Gralla, M.~B., Hajian, A., Halpern, M., Hincks, A.~D., Hlozek, R.,
  Hughes, J.~P., Infante, L., Irwin, K.~D., Kosowsky, A., {Danica Marsden},
  Menanteau, F., Moodley, K., Niemack, M.~D., Nolta, M.~R., Page, L.~A., {Bruce
  Partridge}, Reese, E.~D., Schmitt, B.~L., Sehgal, N., Sherwin, B.~D.,
  Sievers, J., {CristÃ³bal SifÃ³n}, Spergel, D.~N., Staggs, S.~T., Swetz,
  D.~S., Switzer, E.~R., Thornton, R., {Hy Trac}, and Wollack, E.~J., ``The
  {Atacama} {Cosmology} {Telescope}: {Sunyaev}-{Zel}'dovich selected galaxy
  clusters at 148 {GHz} from three seasons of data,'' {\em Journal of Cosmology
  and Astroparticle Physics}~{\bf 2013}(07),  008 (2013).

\bibitem{hand_evidence_2012}
Hand, N., Addison, G.~E., Aubourg, E., Battaglia, N., Battistelli, E.~S.,
  Bizyaev, D., Bond, J.~R., Brewington, H., Brinkmann, J., Brown, B.~R., Das,
  S., Dawson, K.~S., Devlin, M.~J., Dunkley, J., Dunner, R., Eisenstein, D.~J.,
  Fowler, J.~W., Gralla, M.~B., Hajian, A., Halpern, M., Hilton, M., Hincks,
  A.~D., Hlozek, R., Hughes, J.~P., Infante, L., Irwin, K.~D., Kosowsky, A.,
  Lin, Y.-T., Malanushenko, E., Malanushenko, V., Marriage, T.~A., Marsden, D.,
  Menanteau, F., Moodley, K., Niemack, M.~D., Nolta, M.~R., Oravetz, D., Page,
  L.~A., Palanque-Delabrouille, N., Pan, K., Reese, E.~D., Schlegel, D.~J.,
  Schneider, D.~P., Sehgal, N., Shelden, A., Sievers, J., SifÃ³n, C.,
  Simmons, A., Snedden, S., Spergel, D.~N., Staggs, S.~T., Swetz, D.~S.,
  Switzer, E.~R., Trac, H., Weaver, B.~A., Wollack, E.~J., Yeche, C., and
  Zunckel, C., ``Evidence of {Galaxy} {Cluster} {Motions} with the {Kinematic}
  {Sunyaev}-{Zel}'dovich {Effect},'' {\em Physical Review Letters}~{\bf 109},
  041101 (July 2012).

\bibitem{parshley_ccat-prime:_2018}
Parshley, S.~C., Kronshage, J., Blair, J., Herter, T., Nolta, M., Stacey,
  G.~J., Bazarko, A., Bertoldi, F., Bustos, R., Campbell, D.~B., Chapman, S.,
  Cothard, N., Devlin, M., Erler, J., Fich, M., Gallardo, P.~A., Giovanelli,
  R., Graf, U., Gramke, S., Haynes, M.~P., Hills, R., Limon, M., Mangum, J.~G.,
  McMahon, J., Niemack, M.~D., Nikola, T., Omlor, M., Riechers, D.~A., Steeger,
  K., Stutzki, J., and Vavagiakis, E.~M., ``{CCAT}-prime: a novel telescope for
  submillimeter astronomy,'' {\em arXiv:1807.06675 [astro-ph]} ,  220 (July
  2018).
\newblock arXiv: 1807.06675.

\bibitem{Galitzki2018}
Galitzki, N., Ali, A., Arnold, K., Ashton, P.~C., Austermann, Baccigalupi, C.,
  Baildon, T., Barron, D., E., J., Beall, J.~A., Beckman, S., Bruno, S.~M.~M.,
  Bryan, S., Calisse, P.~G., Chesmore, G.~E., Chinone, Y., Choi, S., Coppi, G.,
  Crowley, K.~D., Crowley, K.~T., Cukierman, A., Devlin, M.~J., Dicker, S.,
  Dober, B., Duff, S.~M., Dunkley, J., Fabbian, G., Gallardo, P.~A., Gerbino,
  M., Goeckner-Wald, N., Golec, J.~E., Gudmundsson, J.~E., Healy, E.,
  Henderson, S., Hill, C.~A., Hilton, G.~C., Ho, S.~P., Howe, L.~A., Hubmayr,
  J., Jeong, O., Keating, B., Koopman, B.~J., Kuichi, K., Kusaka, A., Lashner,
  J., Lee, A.~T., Li, Y., Limon, M., Lungu, M., Matsuda, F., Mauskopf, P.~D.,
  May, A.~J., McCallum, N., McMahon, J., Nati, F., Niemack, M.~D.,
  Orlowski-Scherer, J.~L., Parshley, S.~C., Piccirillo, L., Rao, M.~S., Raum,
  C., Salatino, M., Seibert, J.~S., Sierra, C., Silva-Feaver, M., Simon, S.~M.,
  Staggs, S.~T., Stevens, J.~R., Suzuki, A., Teply, G., Thornton, R., Tsai, C.,
  Ullom, J.~N., Vavagiakis, E.~M., Vissers, M.~R., Westbrook, B., Wollack,
  E.~J., Xu, Z., and Zhu, N., ``{The Simons Observatory: Instrument
  overview},'' in [{\em Millimeter, Submillimeter, and Far-Infrared Detectors
  and Instrumentation for Astronomy IX}{\nolinebreak\hspace{0.1em}]},  {\em
  \procspie},  10708--3 (in Press 2018).

\bibitem{Dicker18}
Dicker, S.~R., Gallardo, P.~A., Gudmudsson, J.~E., Mauskopf, P.~D., Ali, A.,
  Ashton, P.~C., Coppi, G., Devlin, M.~J., Galitzki, N., Ho, S.~P., Hill,
  C.~A., Hubmayr, J., Keating, B., Lee, A.~T., Limon, M., Matsuda, F., McMahon,
  J., Niemack, M.~D., Orlowski-Scherer, J.~L., Piccirillo, L., Salatino, M.,
  Simon, S.~M., Staggs, S.~T., Thornton, R., Ullom, J.~N., Vavagiakis, E.~M.,
  Wollack, E.~J., Xu, Z., and Zhu, N., ``Cold optical design for the large
  aperture {Simons}' {Observatory} telescope,'' in [{\em Ground-based and
  {Airborne} {Telescopes} {VII}}{\nolinebreak\hspace{0.1em}]},   {\bf 10700},
  107003E, International Society for Optics and Photonics (July 2018).

\bibitem{crowley18}
Crowley, K.~T., Simon, S.~M., Silva-Feaver, M., Goeckner-Wald, N., Ali, A.,
  Austermann, J., Brown, M.~L., Chinone, Y., Cukierman, A., Dober, B., Duff,
  S.~M., Dunkley, J., Errard, J., Fabbian, G., Gallardo, P.~A., Ho, S.-P.~P.,
  Hubmayr, J., Keating, B., Kusaka, A., McCallum, N., McMahon, J., Nati, F.,
  Niemack, M.~D., Puglisi, G., Rao, M.~S., Reichardt, C.~L., Salatino, M.,
  Siritanasak, P., Staggs, S., Suzuki, A., Teply, G., Thomas, D.~B., Ullom,
  J.~N., VergÃ¨s, C., Vissers, M.~R., Westbrook, B., Wollack, E.~J., Xu, Z.,
  and Zhu, N., ``Studies of systematic uncertainties for {Simons}
  {Observatory}: detector array effects,'' in [{\em Millimeter,
  {Submillimeter}, and {Far}-{Infrared} {Detectors} and {Instrumentation} for
  {Astronomy} {IX}}{\nolinebreak\hspace{0.1em}]},   {\bf 10708},  107083Z,
  International Society for Optics and Photonics (July 2018).

\bibitem{salatino18}
Salatino, M., Lashner, J., Gerbino, M., Simon, S.~M., Didier, J., Ali, A.,
  Ashton, P.~C., Bryan, S., Chinone, Y., Coughlin, K., Crowley, K.~T., Fabbian,
  G., Galitzki, N., Goeckner-Wald, N., Gudmundsson, J.~E., Hill, C.~A.,
  Keating, B., Kusaka, A., Lee, A.~T., McMahon, J., Miller, A.~D., Puglisi, G.,
  Reichardt, C.~L., Teply, G., Xu, Z., and Zhu, N., ``Studies of systematic
  uncertainties for {Simons} {Observatory}: polarization modulator related
  effects,'' in [{\em Millimeter, {Submillimeter}, and {Far}-{Infrared}
  {Detectors} and {Instrumentation} for {Astronomy}
  {IX}}{\nolinebreak\hspace{0.1em}]},   {\bf 10708},  1070848, International
  Society for Optics and Photonics (July 2018).

\bibitem{bryan18}
Bryan, S., Simon, S., and {Teply, G. for the Simons Observatory Collaboration},
  ``{Development of Calibration Strategies for the Simons Observatory},'' {\em
  {Submitted to Proc. SPIE}}  (2018).

\bibitem{Zhu2018}
{Zhu}, N., {Orlowski-Scherer}, J.~L., {Xu}, Z., {Ali}, A., {Arnold}, K.~S.,
  {Ashton}, P.~C., {Coppi}, G., {Devlin}, M.~J., {Dicker}, S., {Galitzki}, N.,
  {Gallardo}, P.~A., {Henderson}, S.~W., {Ho}, S.~P., {Hubmayr}, J., {Keating},
  B., T., L.~A., {Limon}, M., {Lungu}, M., {May}, A.~J., {McMahon}, J.,
  {Niemack}, M.~D., {Piccirillo}, L., {Puglisi}, G., {Rao}, M.~S., {Salatino},
  M., {Silva-Feaver}, M., {Simon}, S.~M., {Staggs}, S., {Thornton}, R.,
  {Ullom}, J.~N., {Vavagiakis}, E.~M., {Westbrook}, B., and {Wollack}, E.~J.,
  ``{Simons Observatory large aperture telescope receiver design overview},''
  in [{\em Millimeter, Submillimeter, and Far-Infrared Detectors and
  Instrumentation for Astronomy IX}{\nolinebreak\hspace{0.1em}]},  {\em
  \procspie},  10708--79 (in Press 2018).

\bibitem{zemax_llc_zemax_2016}
Zemax, ``{Getting} {Started} {With} {OpticStudio} 16,'' (Apr. 2016).

\bibitem{GRASP}
``Ticra grasp.'' \url{https://www.ticra.com/}.
\newblock Accessed: 2018/07/05.

\bibitem{Shimon2008}
{Shimon}, M., {Keating}, B., {Ponthieu}, N., and {Hivon}, E., ``{CMB
  polarization systematics due to beam asymmetry: Impact on inflationary
  science},'' {\em \prd}~{\bf 77},  083003 (Apr. 2008).

\bibitem{hecht2002optics}
Hecht, E., ``Optics, 4th edition,'' {\em International edition, Addison-Wesley,
  San Francisco}~{\bf 3},  2 (2002).

\bibitem{Tran2008}
Tran, H., Lee, A., Hanany, S., Milligan, M., and Renbarger, T., ``Comparison of
  the crossed and the gregorian mizuguchi-dragone for wide-field
  millimeter-wave astronomy,'' {\em Appl. Opt.}~{\bf 47},  103--109 (Jan 2008).

\bibitem{cortiglioni_linear_1994}
Cortiglioni, S., ``Linear polarization and the effects of metal reflectors used
  to redirect the beam in microwave radiometers,'' {\em Review of Scientific
  Instruments}~{\bf 65},  2667--2671 (Aug. 1994).

\bibitem{renbarger_measurements_1998}
Renbarger, T., Dotson, J.~L., and Novak, G., ``Measurements of submillimeter
  polarization induced by oblique reflection from aluminum alloy,'' {\em
  Applied Optics}~{\bf 37},  6643--6647 (Oct. 1998).

\bibitem{koopman_optical_2016}
Koopman, B., Austermann, J., Cho, H.-M., Coughlin, K.~P., Duff, S.~M.,
  Gallardo, P.~A., Hasselfield, M., Henderson, S.~W., Ho, S.-P.~P., Hubmayr,
  J., Irwin, K.~D., Li, D., McMahon, J., Nati, F., Niemack, M.~D., Newburgh,
  L., Page, L.~A., Salatino, M., Schillaci, A., Schmitt, B.~L., Simon, S.~M.,
  Vavagiakis, E.~M., Ward, J.~T., and Wollack, E.~J., ``Optical modeling and
  polarization calibration for {CMB} measurements with {ACTPol} and {Advanced}
  {ACTPol},'' in [{\em Millimeter, {Submillimeter}, and {Far}-{Infrared}
  {Detectors} and {Instrumentation} for {Astronomy}
  {VIII}}{\nolinebreak\hspace{0.1em}]},   {\bf 9914},  99142T, International
  Society for Optics and Photonics (July 2016).

\bibitem{barkats_cosmic_2005}
Barkats, D., Bischoff, C., Farese, P., Gaier, T., Gundersen, J.~O., Hedman,
  M.~M., Hyatt, L., {McMahon}, J.~J., Samtleben, D., Staggs, S.~T., Stefanescu,
  E., Vanderlinde, K., and Winstein, B., ``Cosmic microwave background
  polarimetry using correlation receivers with the {PIQUE} and {CAPMAP}
  experiments,'' ~{\bf 159}(1),  1.

\bibitem{codev}
``Synopsys code v.'' \url{https://www.synopsys.com/}.
\newblock Accessed: 2018/07/05.

\bibitem{koopman_spie_2016}
{Koopman}, B., {Austermann}, J., {Cho}, H.-M., {Coughlin}, K.~P., {Duff},
  S.~M., {Gallardo}, P.~A., {Hasselfield}, M., {Henderson}, S.~W., {Ho},
  S.-P.~P., {Hubmayr}, J., {Irwin}, K.~D., {Li}, D., {McMahon}, J., {Nati}, F.,
  {Niemack}, M.~D., {Newburgh}, L., {Page}, L.~A., {Salatino}, M., {Schillaci},
  A., {Schmitt}, B.~L., {Simon}, S.~M., {Vavagiakis}, E.~M., {Ward}, J.~T., and
  {Wollack}, E.~J., ``{Optical modeling and polarization calibration for CMB
  measurements with ACTPol and Advanced ACTPol},'' in [{\em Millimeter,
  Submillimeter, and Far-Infrared Detectors and Instrumentation for Astronomy
  VIII}{\nolinebreak\hspace{0.1em}]},  {\em \procspie} {\bf 9914},  99142T
  (July 2016).

\bibitem{parshley_optical_2018}
Parshley, S.~C., Niemack, M., Hills, R., Dicker, S.~R., DÃ¼nner, R., Erler,
  J., Gallardo, P.~A., Gudmundsson, J.~E., Herter, T., Koopman, B.~J., Limon,
  M., Matsuda, F.~T., Mauskopf, P., Riechers, D.~A., Stacey, G.~J., and
  Vavagiakis, E.~M., ``The optical design of the six-meter {CCAT}-prime and
  {Simons} {Observatory} telescopes,'' in [{\em Ground-based and {Airborne}
  {Telescopes} {VII}}{\nolinebreak\hspace{0.1em}]},   {\bf 10700},  1070041,
  International Society for Optics and Photonics (July 2018).

\bibitem{polarbear17}
{The POLARBEAR Collaboration}, ``A measurement of the cosmic microwave
  background b -mode polarization power spectrum at subdegree scales from two
  years of polarbear data,'' {\em The Astrophysical Journal}~{\bf 848}(2),  121
  (2017).

\bibitem{louis_atacama_2017}
Louis, T., Grace, E., Hasselfield, M., Lungu, M., Maurin, L., Addison, G.~E.,
  Ade, P. A.~R., Aiola, S., Allison, R., Amiri, M., Angile, E., Battaglia, N.,
  Beall, J.~A., de~Bernardis, F., Bond, J.~R., Britton, J., Calabrese, E., Cho,
  H.-m., Choi, S.~K., Coughlin, K., Crichton, D., Crowley, K., Datta, R.,
  Devlin, M.~J., Dicker, S.~R., Dunkley, J., Dünner, R., Ferraro, S., Fox,
  A.~E., Gallardo, P., Gralla, M., Halpern, M., Henderson, S., Hill, J.~C.,
  Hilton, G.~C., Hilton, M., Hincks, A.~D., Hlozek, R., Ho, S. P.~P., Huang,
  Z., Hubmayr, J., Huffenberger, K.~M., Hughes, J.~P., Infante, L., Irwin, K.,
  Kasanda, S.~M., Klein, J., Koopman, B., Kosowsky, A., Li, D., Madhavacheril,
  M., Marriage, T.~A., McMahon, J., Menanteau, F., Moodley, K., Munson, C.,
  Naess, S., Nati, F., Newburgh, L., Nibarger, J., Niemack, M.~D., Nolta,
  M.~R., Nuñez, C., Page, L.~A., Pappas, C., Partridge, B., Rojas, F., Schaan,
  E., Schmitt, B.~L., Sehgal, N., Sherwin, B.~D., Sievers, J., Simon, S.,
  Spergel, D.~N., Staggs, S.~T., Switzer, E.~R., Thornton, R., Trac, H., Treu,
  J., Tucker, C., Van~Engelen, A., Ward, J.~T., and Wollack, E.~J., ``The
  {Atacama} {Cosmology} {Telescope}: {Two}-{Season} {ACTPol} {Spectra} and
  {Parameters},'' {\em Journal of Cosmology and Astroparticle Physics}~{\bf
  2017},  031--031 (June 2017).
\newblock arXiv: 1610.02360.

\bibitem{fluxa_rojas_far_2016}
Fluxa~Rojas, P.~A., Dünner, R., Maurin, L., Choi, S.~K., Devlin, M.~J.,
  Gallardo, P.~A., Ho, S.-P.~P., Koopman, B.~J., Louis, T., McMahon, J.~J.,
  Nati, F., Niemack, M.~D., Newburgh, L., Page, L.~A., Salatino, M., Schillaci,
  A., Schmitt, B.~L., Simon, S.~M., Staggs, S.~T., and Wollack, E.~J., ``Far
  sidelobe effects from panel gaps of the {Atacama} {Cosmology} {Telescope},''
  in [{\em Millimeter, {Submillimeter}, and {Far}-{Infrared} {Detectors} and
  {Instrumentation} for {Astronomy} {VIII}}{\nolinebreak\hspace{0.1em}]},
  {\bf 9914},  99142Q, International Society for Optics and Photonics (July
  2016).

\bibitem{gallardo18}
Gallardo, P. and Cothard, N.~P., ``{Far Sidelobes from Baffles and Telescope
  Support Structures in the Atacama Cosmology Telescope},'' {\em {Submitted to
  Proc. SPIE}}  (2018).

\bibitem{datta_large-aperture_2013}
Datta, R., Munson, C.~D., Niemack, M.~D., McMahon, J.~J., Britton, J., Wollack,
  E.~J., Beall, J., Devlin, M.~J., Fowler, J., Gallardo, P., Hubmayr, J.,
  Irwin, K., Newburgh, L., Nibarger, J.~P., Page, L., Quijada, M.~A., Schmitt,
  B.~L., Staggs, S.~T., Thornton, R., and Zhang, L., ``Large-aperture
  wide-bandwidth antireflection-coated silicon lenses for millimeter
  wavelengths,'' {\em Applied Optics}~{\bf 52},  8747--8758 (Dec. 2013).

\bibitem{hill_bolocalc:_2018}
Hill, C.~A., Bruno, S. M.~M., Simon, S.~M., Ali, A., Arnold, K.~S., Ashton,
  P.~C., Barron, D., Bryan, S., Chinone, Y., Coppi, G., Crowley, K.~T.,
  Cukierman, A., Dicker, S., Dunkley, J., Fabbian, G., Galitzki, N., Gallardo,
  P.~A., Gudmundsson, J.~E., Hubmayr, J., Keating, B., Kusaka, A., Lee, A.~T.,
  Matsuda, F., Mauskopf, P.~D., McMahon, J., Niemack, M.~D., Puglisi, G., Rao,
  M.~S., Salatino, M., Sierra, C., Staggs, S., Suzuki, A., Teply, G., Ullom,
  J.~N., Westbrook, B., Xu, Z., and Zhu, N., ``{BoloCalc}: a sensitivity
  calculator for the design of {Simons} {Observatory},'' {\em arXiv:1806.04316
  [astro-ph]}  (June 2018).
\newblock arXiv: 1806.04316.

\bibitem{stevens_designs_2018}
Stevens, J.~R., Goeckner-Wald, N., Keskitalo, R., McCallum, N., Ali, A.,
  Borrill, J., Brown, M.~L., Chinone, Y., Gallardo, P.~A., Kusaka, A., Lee,
  A.~T., McMahon, J., Niemack, M.~D., Page, L., Puglisi, G., Salatino, M., Mak,
  S. Y.~D., Teply, G., Thomas, D.~B., Vavagiakis, E.~M., Wollack, E.~J., Xu,
  Z., and Zhu, N., ``Designs for next generation {CMB} survey strategies from
  {Chile},'' in [{\em Millimeter, {Submillimeter}, and {Far}-{Infrared}
  {Detectors} and {Instrumentation} for {Astronomy}
  {IX}}{\nolinebreak\hspace{0.1em}]},   {\bf 10708},  1070841, International
  Society for Optics and Photonics (July 2018).

\bibitem{wollack_measurement_1993}
Wollack, E.~J., Jarosik, N.~C., Netterfield, C.~B., Page, L.~A., and Wilkinson,
  D., ``A {Measurement} of the {Anisotropy} in the {Cosmic} {Microwave}
  {Background} {Radiation} at {Degree} {Angular} {Scales},'' {\em The
  Astrophysical Journal}~{\bf 419},  L49 (1993).

\bibitem{wollack_instrument_1997}
Wollack, E.~J., Devlin, M.~J., Jarosik, N., Netterfield, C.~B., Page, L., and
  Wilkinson, D., ``An {Instrument} for {Investigation} of the {Cosmic}
  {Microwave} {Background} {Radiation} at {Intermediate} {Angular} {Scales},''
  {\em The Astrophysical Journal}~{\bf 476}(2),  440 (1997).

\bibitem{bryan2018development}
Bryan, S.~A., Simon, S.~M., Gerbino, M., Teply, G., Ali, A., Chinone, Y.,
  Crowley, K., Fabbian, G., Gallardo, P.~A., Goeckner-Wald, N., Keating, B.,
  Koopman, B., Kusaka, A., Matsuda, F., Mauskopf, P., McMahon, J., Nati, F.,
  Puglisi, G., Reichardt, C.~L., Salatino, M., Xu, Z., and Zhu, N.,
  ``Development of calibration strategies for the {Simons} {Observatory},'' in
  [{\em Millimeter, {Submillimeter}, and {Far}-{Infrared} {Detectors} and
  {Instrumentation} for {Astronomy} {IX}}{\nolinebreak\hspace{0.1em}]},   {\bf
  10708},  1070840, International Society for Optics and Photonics (July 2018).

\end{thebibliography}
\bibliographystyle{spiebib}

\end{document}